\documentclass[11pt]{article}
\usepackage{jheppub}
\usepackage{amsmath, amsfonts, amssymb, array}
\usepackage{mathrsfs}
\usepackage{slashed}
\newcolumntype{C}[1]{>{\centering\let\newline\\\arraybackslash\hspace{0pt}}m{#1}}
\makeatletter
\g@addto@macro\bfseries{\boldmath}
\makeatother
\title{New higher derivative action for tensor multiplet in $\mathcal{N} = 2$ conformal supergravity in four dimensions}

\author{Subramanya Hegde and Bindusar Sahoo}

\affiliation{
Indian Institute of Science Education and Research,
Vithura, Thiruvananthapuram - 695551, India}

\emailAdd{smhegde14@iisertvm.ac.in}
\emailAdd{bsahoo@iisertvm.ac.in}

\abstract{We will use the covariant superform approach to develop a new density formula for $\mathcal{N}=2$ conformal supergravity which is based on a fermionic multiplet whose lowest component is a dimension-5/2 spinor. We will show that this density formula admits an embedding of the real scalar multiplet of \cite{Hegde:2017sgl}. Upon using the embedding of the tensor multiplet into the real scalar multiplet, we will construct a new higher derivative action of the tensor multiplet in $\mathcal{N}=2$ conformal supergravity.} 

\begin{document}
\allowdisplaybreaks
\maketitle
\section{Introduction}
Supergravity theories are supersymmetric extensions of gravity which, apart from diffeomorphism, are also invariant under local supersymmetry and $\textit{possibly}$ some other gauge symmetries.  Although they may not be ultraviolet complete on their own, they do describe the low energy sector of string theory which is one of the candidates for a complete quantum theory of gravity. String theory compactified on different compact manifolds gives rise to effective supergravity theories in the low energy limit. Construction of supergravity invariants, therefore, is crucial to understand leading order and higher derivative corrections to black hole entropy arising in the context of string theory. For a class of extremal black holes, such higher derivative corrections have been well studied, see e.g. \cite{LopesCardoso:1998tkj,Sen:2007qy,Mohaupt:2000mj}, facilitated by the construction of various higher derivative invariants in $\mathcal{N}=2$ supergravity in four dimensions\cite{Bergshoeff:1980is,Butter:2013lta,Kuzenko:2015jxa,deWit:2010za}. 

While supergravity invariants second order in derivatives can be constructed with on-shell supersymmetry techniques, to construct higher derivative or more general matter couplings in supergravity, off-shell methods are found to be useful. This is due to the fact that modifying the on-shell supersymmetry transformations to construct higher derivative actions or general matter couplings, involves the arduous task of iteratively modifying the supersymmetry rules and the equations of motion such that the supersymmetry algebra is satisfied on-shell with respect to the modified equations of motion. In off-shell supersymmetry, the supersymmetry algebra is satisfied off-shell and the transformations are independent of equations of motion which makes them suitable for construction of general matter couplings and higher derivative invariants in supergravity.

Conformal supergravity, in particular superconformal ``multiplet calculus'', provides a powerful method to construct supergravity invariants using off-shell supersymmetry. Here, the large number of symmetries in the superconformal algebra allow for shorter representations than for Poincar\'e supergravity. Physical Poincar\'e supergravity is then obtained by use of compensator fields to gauge fix the additional symmetries in the superconformal theory \cite{Kaku:1978ea}. Upon gauge fixing, actions for matter multiplets in conformal supergravity lead to kinetic actions for gravity multiplet in Poincar\'e supergravity.

The multiplet that contains all the gauge fields of the superconformal algebra is known as the Weyl multiplet. In $\mathcal{N}=2$ conformal supergravity in four dimensions, the transformation of the Weyl multiplet and the complete action was constructed in \cite{Ferrara:1977ij, deWit:1979dzm, Bergshoeff:1980is}. The Weyl multiplet contains all the gauge fields of the superconformal algebra along with few auxiliary fields required for the closure of the multiplet. It has $24+24$ (bosonic+fermioinic) off-shell degrees of freedom. Different choices of the auxiliary fields lead to different types of Weyl multiplet. The Weyl multiplet discussed in \cite{Bergshoeff:1980is} is known as the standard Weyl multiplet. A different set of auxiliary fields lead to the dilaton Weyl multiplet in four dimensions \cite{Butter:2017pbp}. We will use the standard Weyl multiplet in this work. All the details of the $\mathcal{N}=2$ standard Weyl multiplet is discussed in appendix-\ref{N2Weyl}. Apart from the Weyl multiplet, there can be several matter multiplets in $\mathcal{N}=2$ conformal supergravity. Transformation rules for the matter multiplets such as the 8+8 components vector multiplet and 8+8 components tensor multiplet were presented in \cite{deWit:1979dzm}. A larger matter multiplet containing 24+24 off-shell degrees of freedom called the real scalar multiplet was discussed in \cite{Hegde:2017sgl} which was the generalization of the flat space multiplet discussed in \cite{Howe:1982tm} to conformal supergravity. We will need the tensor multiplet and the real scalar multiplet for our work. The components of the tensor multiplet and its supersymmetry transformation rules will be discussed in appendix-\ref{N2Weyl} and the details about real scalar multiplet will be discussed in section-\ref{Real}.

A crucial ingredient to construct actions for these matter multiplets, and thereby Poincar\'e supergravity theories is the $\mathcal{N}=2$ chiral density formula \cite{deRoo:1980mm}. The term density formula refers to a set of objects that appear in the action accompanied by fields such as the vielbein and gravitino. For the action to be invariant under supersymmetry, they transform among each other in a particular way, and satisfy certain constraints such as chirality or reality. 

The chiral density formula of \cite{deRoo:1980mm} was based on a $16+16$ components chiral multiplet. It was shown that this $16+16$ multiplet can be reduced to an $8+8$ restricted chiral multiplet upon introducing a consistent set of constraints. Further, an embedding was found for the gauge invariant objects of the $8+8$ vector multiplet inside this restricted chiral multiplet. This allowed for the construction of a superconformal action for the vector multiplet in conformal supergravity background and thereby resulted in the construction of a minimal Poincar\'e supergravity theory upon gauge fixing \cite{deWit:1980lyi}. While the free tensor gauge field action is not conformal in four dimensions, an improved tensor multiplet action was constructed, based on a density formula made out of the components of a linear multiplet\footnote{The components of Linear multiplet are the gauge invariant objects of a tensor multiplet}  and a vector muiltiplet, in \cite{deWit:1982na} which allowed for a new minimal formulation of Poincar\'e supergravity.  

Construction of density formula such as the chiral density formula and the linear-vector density formula played a key role in the construction of supergravity theories. The process of embedding products of different multiplets into density formulae is known as the superconformal multiplet calculus. 

The three minimal Poincar\'e supergravity theories constructed by this method, are based on the actions of two compensating multiplets in the background of Weyl multiplet. Vector multiplet compensator is necessary to obtain the graviphoton in Poincar\'e supergravity. However this alone is insufficient, as the auxiliary field $D$ of the Weyl multiplet occurs linearly in the action and its equation of motion places a non trivial condition on the fields that is inconsistent with the dilatation gauge condition. To resolve this, another compensating multiplet is needed. Tensor multiplet or nonlinear multiplet, when used for this purpose lead to off-shell Poincar\'e supergravity and when the on-shell hypermultiplet is used, one obtains an on-shell Poincar\'e supergravity.

The procedure of superconformal multiplet calculus has been used in the past to construct several higher derivative invariants, in various dimensions and of relevance to us here, in four dimensions \cite{Bergshoeff:1980is,Butter:2013lta,Kuzenko:2015jxa,deWit:2010za}. Recently, higher derivative couplings have also been discussed in the literature in the projective superspace formulation of conformal supergravity \cite{Kuzenko:2009zu, Butter:2010jm, Butter:2012sb}. These results have played an important role in understanding higher derivative corrections to black hole entropy. Thus construction of new density formula, thereby new Poincar\'e supergravity invariants is of interest, as it allows us to study and understand these results further.       

In this paper, we construct a new density formula in four dimensional $\mathcal{N}=2$ conformal supergravity, using the covariant superform method. This is a standard method which is used to construct invariant actions in supergravity in various dimensions (See \cite{Butter:2012ze, Butter:2013rba, Kuzenko:2013vha, Butter:2014xxa}). It was first introduced in superspace as the `ectoplasm principle' \cite{Gates:1997ag,Gates:1997kr} and has been recently applied in \cite{Butter:2019edc} to construct the most general actions for $\mathcal{N}=4$ conformal supergravity. Similar ideas have also been used in rheonomic approach to supergravity \cite{DAuria:1982mkx,Castellani:1991eu}.  We will elaborate on this method later. The crucial difference of the new density formula from the earlier ones is that the lowest component (i.e the component with the lowest Weyl weight) of the multiplet from which it is constructed is a spinor $\Sigma_{ijk}$. It is a superconformal primary field\footnote{A field that is invariant under special conformal transformation and S-supersymmetry is known as superconformal primary field. S-supersymmetry is a different kind of supersymmetry that arises in conformal supergravity. It is different from the ordinary supersymmetry which is referred to as the Q-supersymmetry. For details we refer the reader to appendix-\ref{N2Weyl}}, transforms in the $\textbf{4}$ of SU(2) R symmetry, has chirality $-1$, chiral weight $-1/2$ and Weyl weight $+5/2$. Along with its supersymmetry descendents, it appears in the density formula as:
\begin{align}
S=\int \mathcal{L} d^4 x \;,
\end{align}
where the Lagrangian density $\mathcal{L}$ is given as:
\begin{align}\label{d4.8}
e^{-1}\mathcal{L}&=-i\mathcal{L}+2i\bar{\psi}_{ai}\Upsilon^{a}_{j}\varepsilon^{ij}+2i\bar{\psi}_{ai}\gamma^{a}\Psi_{j}\varepsilon^{ij}+\frac{3i}{16}\bar{\psi}_{a}^{j}\gamma^{ab}\gamma^{c}\psi_{bk}\mathcal{G}_{cjl}\varepsilon^{kl}+\frac{i}{16}\bar{\psi}_{a}^{j}\gamma^{c}\gamma^{ab}\psi_{bk}\mathcal{G}_{cjl}\varepsilon^{kl}\nonumber \\
& \;\;\; -\frac{i}{4}\bar{\psi}_{a}^{i}\gamma^{ab}\psi_{b}^{j}\mathcal{D}^{kl}\varepsilon_{ik}\varepsilon_{jl}-\frac{i}{32}\bar{\psi}_{a}^{i}\gamma^{ab}\gamma^{cd}\psi_{b}^{j}\mathcal{B}_{cd}^{kl}\varepsilon_{ik}\varepsilon_{jl}-i\varepsilon^{abcd}\bar{\psi}_{ai}\gamma_{b}\psi_{c}^{j}\bar{\psi}_{dk}\Sigma_{jmn}\varepsilon^{km}\varepsilon^{in}+\text{h.c}\;.
\end{align}
All the components appearing in the above density formula are related to $\Sigma_{ijk}$ by subsequent Q-supersymmetry transformations. As one can also see, the maximum number of gravitno appearing in the above density formula is three as opposed to four in the chiral density formula or two in the linear-vector density formula. Another crucial difference with the other $\mathcal{N}=2$ density forumlae is that, not all components of the multiplet appear in the density formulae. For example, there are components $\mathcal{C}_{ijkl}$ and $\mathcal{H}^{a}_{ijkl}$ that appear in the supersymmetry transformation of $\Sigma_{ijk}$ (\ref{d2.3}, \ref{d3.2}) but do not appear in the above density formula. Analogous density formula has also been found in six dimensional supergravity \cite{Butter:2016qkx, Butter:2017jqu} based on the study of a super six-form which was first studied in flat space in \cite{Arias:2014ona}. The density formula is based on a fermionic multiplet whose lowest component is a dimension 9/2 spinor. It seems plausible, although non trivial, that both the density formulae are related by a dimensional reduction. 

We will also show that the density formula (\ref{d4.8}) embeds the $24+24$ real scalar multiplet \cite{Hegde:2017sgl} which further admits a tensor multiplet embedding. We will use the embedding of the tensor multiplet, to construct a new higher derivative invariant for the tensor multiplet in $\mathcal{N}=2$ conformal supergravity.

The paper is organized as follows. In section-\ref{density} we will construct a new density formula using the covariant superform approach. We will elaborate on the covariant superform approach in the same section. In section-\ref{Real} we will review the real scalar multiplet of \cite{Hegde:2017sgl} along with the tensor multilplet embedding. In section-\ref{Real_action} we will find the invariant action for the real scalar multiplet by embedding it into the density formula constructed in section-\ref{density}. In section-\ref{tensor_action}, we will use tensor multiplet embedding in the real scalar multiplet to obtain new higher derivative coupling of the tensor multiplet in conformal supergravity. We will end with conclusions and future directions.
\section{A new density formula for $\mathcal{N}=2$ conformal supergravity}\label{density}
In this section, we will build a new density formula using the covariant superform approach. We will briefly outline the method and use it to construct a new density formula for $\mathcal{N}=2$ conformal supergravity. In order to understand this method, consider the following action integral of a d-form ($J$) in a d-dimensional submanifold ($M_d$) embedded in a larger D-dimensional manifold ($\mathcal{M}_D$):
\begin{align}\label{density-1}
S=\int_{M_d}J
\end{align}
Under a general diffeomorphism ($\xi$), the d-form transforms as:
\begin{align}\label{density-2}
\delta_{\xi}J=i_{\xi}dJ+d(i_{\xi} J)
\end{align}
If the submanifold ($M_d$) is closed or if there are appropriate fall-off boundary conditions, the second term in the above variation (\ref{density-2}) will not contribute to the variation of the action integral (\ref{density-1}). Hence, the action-integral will be invariant under a general diffeomorphism in the larger Manifold ($\mathcal{M}_D$) if the d-form ($J$) satisfies the following condition:
\begin{align}\label{density-3}
dJ=0
\end{align}
Now consider the case where $M_d$ is the space-time manifold and $\mathcal{M}_D$ is a larger manifold where some of the gauge symmetries (for eg: supersymmetry) has been geometerized. In that case, the corresponding gauge transformations will be a part of the generalized diffeomorphism of the larger manifold and hence the action-integral defined on the space-time manifold will be invariant under the corresponding gauge transformation if it satisfies the closure condition (\ref{density-3}).

For further discussions, let us restrict ourselves to conformal supergravity in four spacetime dimensions. Apart from diffeomorphism, the other gauge-symmetries present in conformal supergravity are local Lorentz transformation, R-symmetry, dilatation, ordinary supersymmetry (also known as Q-supersymmetry) as well as special supersymmetry (also known as S-supersymmetry). For our purpose, we will consider the case where the larger Manifold ($\mathcal{M}$) is a superspace where the Q-supersymmetry has been geometerized. The corresponding fermionic directions are labelled as $\theta^{m}$. The generalized vielbein (or supervielbein) will have legs along the spacetime manifold as well as the fermionic directions as shown below:
\begin{align}\label{density-4}
E^A=dx^{\mu}E_{\mu}^{A}+d\theta^{m}E_{m}^{A}
\end{align}
The supervielbein 1-form when restricted to the spacetime manifold (i.e $\theta=d\theta=0$) are nothing but $(E^A)_{|\theta=d\theta=0}=(e^a,\frac{1}{2}\psi^{i},\frac{1}{2}\psi_{i})$, where $e^a$ is the vielbein 1-form, $\psi^{i}$ and $\psi_{i}$ are the left-chiral and right-chiral gravitino 1-form respectively\footnote{We will be dealing with fermions in the Dirac 4-component notation}. The four form $J$ that we will need for the action integral can be further decomposed as:
\begin{align}\label{density-5}
J=J_{DCBA}E^A E^B E^C E^D\;,
\end{align}
where the wedge product between the 1-forms is assumed and the block $J_{DCBA}$ is fully supercovariant. The four-form is further assumed to be invariant under all the other gauge transformations of conformal supergravity except Q-supersymmetry. Invariance under Q-supersymmetry is guaranteed if the four-form ($J$) satisfies the closure condition (\ref{density-3}). However since we have also assumed that $J$ is invariant under other gauge transformations, we can replace the closure condition with the covariant closure condition:
\begin{align}\label{density-6}
\nabla J=0\;,
\end{align}
where $\nabla$ is an exterior derivative that is covariant w.r.t all the other gauge transformations of conformal supergravity except Q-supersymmetry. The invariance of the four-form $J$ under local Lorentz transformations, dilatation, special conformal transformation and R-symmetry is manifest by taking the term in the Lagrangian corresponding to separate blocks $J_{DCBA}$ to be invariant under these symmetries. Invariance under S-supersymmetry would mean that the different blocks will be related to each other by S-transformation since gravitino transforms to a vielbein under S-transformation as given below:
\begin{align}\label{s-grav}
\delta_{S}\psi_{\mu}{}^{i}=-e_{\mu}^{a}\gamma_a \eta^{i}\;.
\end{align}
While imposing the covariant closure condition, the supervielbein appearing in the four form $J$ (\ref{density-5})will be taken to be the full supervielbein $E^A$ that has legs along the full superspace $\mathcal{M}$ instead of the restricted superielbein $E^A_{|\theta=d\theta=0}$ that has legs only along the spacetime manifold $M$. However, we will somewhat abuse the notation and write $E^A=(e^a, \frac{1}{2}\psi^{i},\frac{1}{2}\psi_{i})$ while using the covariant closure condition (\ref{density-6}). Once we solve the covariant closure condition and get an appropriate 4-form $J$, the supervielbein in the action-integral will only involve the usual vielbein and the gravitino fields since the action integral is defined on the spacetime manifold $M$ which is a $\theta=d\theta=0$ slice of the full manifold $\mathcal{M}$. In order to use the covariant closure condition, we must know how the covariant exterior derivative acts on the supervielbein and supercovariant objects. The covariant exterior derivative acts on the supervielbeins to give the corresponding superspace torsion tensors as shown below\footnote{For a comprehensive treatment of $\mathcal{N}=2$ conformal supergravity in superspace, we refer the reader to \cite{Butter:2011sr}}:
\begin{align}\label{density-7}
\nabla e^{a}&=-\frac{1}{2}\bar{\psi}^{i}\gamma^{a}\psi_{i} \equiv t_{0} e^{a}\;, \nonumber \\
\nabla \psi^{i} &=\frac{1}{2}e^{a}e^{b}R(Q)_{ba}{}^{i}-\frac{1}{16}\gamma\cdot T^{ij}e^{a}\gamma_{a}\psi_{j}\equiv t_{3/2}\psi^{i}+t_{1}\psi^{i}\;, \nonumber \\
\nabla \psi_{i}&=\frac{1}{2}e^{a}e^{b}R(Q)_{ba}{}_{i}-\frac{1}{16}\gamma\cdot T_{ij}e^{a}\gamma_{a}\psi^{j}\equiv \bar{t}_{3/2}\psi_{i}+\bar{t}_{1}\psi_{i}\;,
\end{align}
where, following \cite{Butter:2019edc}, we have introduced shorthands $t_{n}$ and $\bar{t}_{n}$ for the torsion 2-forms. The subscripts denote the Weyl weight of the covariant fields appearing in the expressions. The covariant exterior derivative acts on the super-covariant objects as:
\begin{align}\label{density-8}
\nabla \Phi \equiv (\nabla_{1}+\nabla_{1/2}+\bar{\nabla}_{1/2})\Phi
\end{align}
where,
\begin{align}\label{density-9}
\nabla_{1}\Phi&=e^{a}D_{a}\Phi\;,\nonumber \\
\nabla_{1/2}\Phi&=\frac{1}{2}\bar{\psi}^{k}\nabla_{k}\Phi=\delta^{L}_{Q}\left(\frac{1}{2}\psi\right)\Phi\;, \nonumber \\
\bar{\nabla}_{1/2}\Phi&=\frac{1}{2}\bar{\psi}_{k}\nabla^{k}\Phi =\delta^{R}_{Q}\left(\frac{1}{2}\psi\right)\Phi\;, \nonumber \\
\end{align}
where, $D_{a}\Phi$ is the fully super-covariant derivative of $\Phi$ and $\delta_{Q}^{(L)}\left(\frac{1}{2}\psi\right)\Phi$ or $\delta_{Q}^{(R)}\left(\frac{1}{2}\psi\right)\Phi$ is either the the left or the right Q-supersymmetry transformation of $\Phi$ respectively where the parameter has been replaced by $\frac{1}{2}\psi$. The subscripts on the $\nabla$ in the shorthand notation given in (\ref{density-8}) denotes the Weyl weight of the corresponding operators acting on $\Phi$. With the above definitions at hand, the covariant closure condition on $J$ can be written as:
\begin{align}\label{density-10}
\nabla J=(t_0+t_{1/2}+t_{1}+\bar{t}_{1/2}+\bar{t}_{1}+\nabla_{1}+\nabla_{1/2}+\bar{\nabla}_{1/2})J=0
\end{align}
In the above formalism it is easier to decompose the above covariant closure condition on $J$ by the number of gravitino factors contained in them and set them individually to zero. In order to apply this formalism, one has to start with an ansatz which is basically the structure of the term in $J$ that contains the maximum number of gravitino. For example one may start with the following ansatz for the structure of the highest gravitino term in $J$:
\begin{align}\label{chiral}
J_{\bar{\psi}^4}=\bar{\psi}_{i}\psi_{j}\bar{\psi}_{k}\psi_{l}\varepsilon^{ij}\varepsilon^{kl}A\;,
\end{align}
where $A$ is a Lorentz as well as SU(2) scalar and has the required Weyl weight of +2 and chiral weight of -2. Upon imposing the covariant closure condition, one can recover the well-known chiral density formula along with the full supersymmetry transformations of the chiral multiplet with $A$ as its lowest component. In the above equation, the subscript denote the number of right-handed gravitino it carries. In general we will decompose $J$ as:
\begin{align}\label{density-11}
J=\sum_{m+n+p=4}J_{e^m\psi^n\bar{\psi}^p}\;,
\end{align}
where, $m$ is the number of vielbein, $n$ is the number of left-handed gravitino and $p$ is the number of right-handed gravitino. 

Now, we will take a different route and propose the following structure for the highest gravitino term in $J$:
\begin{align}\label{density-12}
J_{e\bar{\psi}^{2}\psi}&= e^{a}\bar{\psi}_{i}\gamma_{a}\psi^{j}\bar{\psi}_{k}\Sigma_{jmn}\varepsilon^{km}\varepsilon^{in}\;,\nonumber \\
J_{{e\psi}^{2}\bar{\psi}}&=e^{a}\bar{\psi}_{i}\gamma_{a}\psi^{j}\bar{\psi}^{k}\Sigma^{imn}\varepsilon_{km}\varepsilon_{jn}\;,
\end{align}
The component $\Sigma_{ijk}$ appearing above is a spinor field which has chirality $-1$, chiral weight $-1/2$ and Weyl weight $+5/2$. The other object $\Sigma^{ijk}$ that appears above is related to $\Sigma_{ijk}$ by charge conjugation and has the opposite chirality and chiral weight but the same Weyl weight\footnote{We will be following the \textit{chiral notation} whereby raising and lowering of SU(2) indices results in opposite chiral weight and chirality but Weyl weight remains unchanged.}. Both $\Sigma_{ijk}$ and $\Sigma^{ijk}$ are superconformal primary fields and transform in the $\bf{4}$ representation of $SU(2)$-R symmetry. Let us now apply the covariant closure condition $\nabla J=0$ on the above highest gravitino term. As discussed previously, the covariant closure condition, which is a five-form, can be decomposed according to the number of gravitino 1-form it carries as shown below:
\begin{align}\label{density-13}
\nabla J &=\sum_{m+n+p=5}(\nabla J)_{e^m \psi^n \bar{\psi}^{p}}=0\nonumber \\
& \implies (\nabla J)_{e^m\psi^n\bar{\psi}^{p}}=0
\end{align}
The individual $(\nabla J)_{e^m \psi^n \bar{\psi}^p}$ are typically referred to as $e^m \psi^n \bar{\psi}^p$ Bianchi identity. 
\subsection{Solving the $\bar{\psi}^3\psi^2$ Bianchi}
This arises only from $t_{0} J_{e\bar{\psi}^2 \psi}$ and should cancel on its own. As we will show below, it indeed cancels on its own.
\begin{align}\label{d1.1}
t_{0} J_{e\bar{\psi}^{2}\psi}&=\frac{1}{2}\bar{\psi}^{\ell}\gamma^{a}\psi_{\ell}\bar{\psi}_{i}\gamma_{a}\psi^{j}\bar{\psi}_{k}\Sigma_{jmn}\varepsilon^{km}\varepsilon^{in}\nonumber \\
&=\bar{\psi}_{i}\psi_{\ell}\bar{\psi}^{\ell}\psi^{j}\bar{\psi}_{k}\Sigma_{jmn}\varepsilon^{km}\varepsilon^{in}=0
\end{align}
We did a Fierz rearrangement in coming from the first line to the next line. One easy way to see how the last line becomes zero is to understand that $\bar{\psi}_{i}\psi_{\ell}$ is anti-symmetric in SU(2) indices $i$ and $\ell$ and hence proportional to $\varepsilon_{i\ell}$. Explicitly $\bar{\psi}_{i}\psi_{\ell}=\frac{1}{2}\varepsilon_{i\ell}\varepsilon^{np}\bar{\psi}_{n}\psi_{p}$. Similarly ${\psi}^{\ell}\psi^{j}$ is proportional to $\varepsilon^{\ell j}$. Explicitly $\bar{\psi}^{\ell}\psi^{j}=\frac{1}{2}\varepsilon^{\ell j}\varepsilon_{np}\bar{\psi}^{n}\psi^{p}$. Hence $\bar{\psi}_{i}\psi_{\ell}\bar{\psi}^{\ell}\psi^{j}$ is proportional to $\delta_{i}^{j}$. Since $
\Sigma_{ijk}$ transforms in the $\bf{4}$ irrep of SU(2)-R symmetry, this implies $\delta_{i}^{j}\Sigma_{jmn}\varepsilon^{km}\varepsilon^{in}=0$ and hence the expression in (\ref{d1.1}) goes to zero. Similarly one can show that the conjugate Bianchi $\psi^3\bar{\psi}^2$ is also automatically satisfied.
\subsection{Solving the $e\bar{\psi}^{3}\psi$ Bianchi}
This comes from $\bar{\nabla}_{1/2}J_{e\bar{\psi}^2\psi}$ and $t_{0}J_{e^2\bar{\psi}^{2}}$. We already know what is the structure of $J_{e\bar{\psi}^2\psi}$ from (\ref{density-12}). By solving this Bianchi, our aim is to relate the fields appearing in $J_{e^2\bar{\psi}^{2}}$ to the supersymmetry transformation of $\Sigma_{ijk}$. We will also see that some fields appearing in the supersymmetry transformation of $\Sigma_{ijk}$ will be constrained. In order to solve this Bianchi, let us define:
\begin{align}\label{d2.1}
& (\bar{\nabla}^{\ell}\gamma_{bc}\Sigma_{jmn})_{\bf{5}}=\varepsilon^{\ell p}\mathcal{A}_{bc jmnp}\;, \;\;\; \bar{\nabla}^{p}\gamma_{bc}\Sigma_{mnp}=\mathcal{B}_{bcmn}\;, \nonumber \\
&  (\bar{\nabla}^{\ell}\Sigma_{jmn})_{\bf{5}}=\varepsilon^{\ell p}\mathcal{C}_{jmnp}\;, \;\;\; \bar{\nabla}^{p}\Sigma_{mnp}=\mathcal{D}_{mn}\;.
\end{align}
In terms of the above defined irreps of SU(2)-R symmetry, the full decompositions of the operator $\bar{\nabla}^{\ell}$ acting on $\Sigma_{ijk}$ takes the following form:
\begin{align}\label{d2.2}
\bar{\nabla}^{\ell}\gamma_{bc}\Sigma_{jmn}&=\varepsilon^{\ell p}\mathcal{A}_{bc jmnp}+\frac{3}{4}\delta^{\ell}_{(j}\mathcal{B}_{bcmn)}\;,\nonumber \\
\bar{\nabla}^{\ell}\Sigma_{jmn}&=\varepsilon^{\ell p}\mathcal{C}_{jmnp}+\frac{3}{4}\delta^{\ell}_{(j}\mathcal{D}_{mn)}\;.
\end{align}
And hence, in terms of the fields defined above, the right supersymmetry transformation of $\Sigma_{ijk}$ takes the following form:
\begin{align}\label{d2.3}
\delta_{Q}^{R}\Sigma_{ijk}&=\bar{\epsilon}_{\ell}\nabla^{\ell}\Sigma_{ijk}\nonumber \\
&=-\frac{1}{2}\bar{\nabla}^{\ell}\Sigma_{ijk}\epsilon^{\ell}+\frac{1}{8}\bar{\nabla}^{\ell}\gamma_{ab}\Sigma_{ijk}\gamma^{ab}\epsilon^{\ell}\nonumber \\
&=-\frac{1}{2}\varepsilon^{\ell m}\mathcal{C}_{ijkm}\epsilon_{\ell}-\frac{3}{8}\mathcal{D}_{(ij}\epsilon_{k)}+\frac{1}{8}\varepsilon^{\ell m}\mathcal{A}_{abijkm}\gamma^{ab}\epsilon_{\ell}+\frac{3}{32}\mathcal{B}_{ab(ij}\gamma^{ab}\epsilon_{k)}\;.
\end{align}
In terms of the above fields, $\bar{\nabla}_{1/2}J_{e\bar{\psi}^{2}\psi}$ takes the following form:
\begin{align}\label{d2.4}
\bar{\nabla}_{1/2}J_{e\bar{\psi}^{2}\psi}&=-\frac{1}{8}t_{0}\left(e^{a}e^{b}\right)\bar{\psi}_{i}\gamma_{ab}\psi_{j}\mathcal{D}_{k\ell}\varepsilon^{ik}\varepsilon^{j\ell}\nonumber \\
&\;\;\;+\frac{1}{16}e^{a}\bar{\psi}_{i}\gamma_{a}\psi^{j}\bar{\psi}_{k}\gamma_{bc}\psi_{\ell}\mathcal{A}^{bc}_{jmnp}\varepsilon^{km}\varepsilon^{in}\varepsilon^{\ell p}\nonumber \\
&\;\;\;-\frac{1}{64}t_{0}\left(e^{a}e^{b}\right)\bar{\psi}_{i}\gamma_{ab}\gamma_{cd}\psi_{j}\mathcal{B}^{cd}_{k\ell}\varepsilon^{ik}\varepsilon^{j\ell}\;.
\end{align}
It is clear that the first and third expressions above are $t_0$ exact and hence it will cancel if we take the $t_{0}$ operation on the following $J_{e^2\bar{\psi}^{2}}$:
\begin{align}\label{d2.5}
J_{e^2\bar{\psi}^{2}}&=\frac{1}{8}e^{a}e^{b}\bar{\psi}_{i}\gamma_{ab}\psi_{j}\mathcal{D}_{k\ell}\varepsilon^{ik}\varepsilon^{j\ell}+\frac{1}{64}e^{a}e^{b}\bar{\psi}_{i}\gamma_{ab}\gamma_{cd}\psi_{j}\mathcal{B}^{cd}_{k\ell}\varepsilon^{ik}\varepsilon^{j\ell}\;.
\end{align}
The second term is not $t_0$ exact and will only vanish if we have the following constraint 
\begin{align}\label{d2.6}
\mathcal{A}^{bc}_{jmnp}=0\;.
\end{align}
The conjugate $e{\psi}^{3}\bar{\psi}$ Bianchi will give the constraint $\mathcal{A}_{bc}^{jmnp}=0$, where $\mathcal{A}_{bc}^{jmnp}$ appears in the  left supersymmetry transformation of $\Sigma^{ijk}$ as shown below:
\begin{align}\label{d2.7}
(\bar{\nabla}_{\ell}\gamma_{bc}\Sigma^{jmn})_{\bf{5}}=\varepsilon_{\ell p}\mathcal{A}_{bc}^{jmnp}\;.
\end{align}
However this constraint is automatically satisfied because $\mathcal{A}_{bc}^{jmnp}= (\mathcal{A}_{bcjmnp})^{*}$. The conjugate Bianchi will also give the hermitian conjugate of $J_{e^2\bar{\psi}^{2}}$ which is:
\begin{align}\label{d2.8}
J_{e^2{\psi}^{2}}&=\frac{1}{8}e^{a}e^{b}\bar{\psi}^{i}\gamma_{ab}\psi^{j}\mathcal{D}^{kl}\varepsilon_{ik}\varepsilon_{jl}+\frac{1}{64}e^{a}e^{b}\bar{\psi}^{i}\gamma_{ab}\gamma_{cd}\psi^{j}\mathcal{B}^{cdkl}\varepsilon_{ik}\varepsilon_{jl}\;.
\end{align}
\subsection{Solving the $e\bar{\psi}^{2}\psi^{2}$ Bianchi}
This arises from $\nabla_{1/2}J_{e\bar{\psi}^{2}\psi}$, its hermitian conjugate $\bar{\nabla}_{1/2}J_{e{\psi}^{2}\bar{\psi}}$ and $t_{0}J_{e^2\psi\bar{\psi}}$. By analyzing this Bianchi, we hope to obtain $J_{e^2\psi\bar{\psi}}$ which will contain fields appearing in the left supersymmetry transformation of $\Sigma_{ijk}$. Along with that we will also obtain constraints on some of the fields appearing in the left supersymmetry transformation of $\Sigma_{ijk}$. For this purpose, let us define:
\begin{align}\label{d3.1}
\bar{\nabla}_{(i}\gamma^{b}\Sigma_{jk\ell)}=\mathcal{H}^{b}_{ijk\ell}\;, \;\;\; \varepsilon^{pq}\nabla_{p}\gamma^{b}\Sigma_{qmn}=\mathcal{G}^{b}_{mn}\;.
\end{align}
In terms of the above mentioned fields, the left-supersymmetry transformation of $\Sigma_{ijk}$ becomes:
\begin{align}\label{d3.2}
\delta_{Q}^{L}\Sigma_{ijk}&=\bar{\epsilon}^{\ell}\nabla_{\ell}\Sigma_{ijk}\nonumber \\
&=-\frac{1}{2}\bar{\nabla}_{\ell}\gamma^{b}\Sigma_{ijk}\gamma_{b}\epsilon^{\ell}\nonumber \\
&=-\frac{1}{2}\mathcal{H}^{b}_{ijk\ell}\gamma_{b}\epsilon^{\ell}-\frac{3}{8}\varepsilon_{\ell(i}\mathcal{G}^{b}_{jk)}\gamma_{b}\epsilon^{\ell}\;.
\end{align}
And we get:
\begin{align}\label{d3.3}
\nabla_{1/2}J_{e\bar{\psi}^2\psi}&=-\frac{1}{4}e^{a}\bar{\psi}_{i}\gamma_{a}\psi^{j}\bar{\psi}_{k}\gamma_{b}\psi^{\ell}\mathcal{H}^{b}_{\ell jmn}\varepsilon^{km}\varepsilon^{in}\nonumber \\
&\;\;\; +\frac{3}{32}t_{0}(e^{a}e^{b})\bar{\psi}^{j}\gamma_{ab}\gamma_{c}\psi_{k}\mathcal{G}^{c}_{j\ell}\varepsilon^{k\ell}\nonumber \\
&\;\;\; -\frac{1}{32}t_{0}(e^{a}e^{b})\bar{\psi}^{j}\gamma_{c}\gamma_{ab}\psi_{k}\mathcal{G}^{c}_{j\ell}\varepsilon^{k\ell}\;.
\end{align}
The Hermitian conjugate of the above expression is:
\begin{align}\label{d3.4}
\bar{\nabla}_{1/2}J_{e{\psi}^2\bar{\psi}}&=-\frac{1}{4}e^{a}\bar{\psi}_{j}\gamma_{a}\psi^{i}\bar{\psi}_{\ell}\gamma_{b}\psi^{k}\mathcal{H}^{b}{}^{\ell jmn}\varepsilon_{km}\varepsilon_{in}\nonumber \\
&\;\;\; -\frac{3}{32}t_{0}(e^{a}e^{b})\bar{\psi}^{k}\gamma_{c}\gamma_{ab}\psi_{j}\mathcal{G}^{c}{}^{j\ell}\varepsilon_{k\ell}\nonumber \\
&\;\;\; +\frac{1}{32}t_{0}(e^{a}e^{b})\bar{\psi}^{k}\gamma_{ab}\gamma_{c}\psi_{j}\mathcal{G}^{c}{}^{j\ell}\varepsilon_{k\ell}\;.
\end{align}
The first expressions in the above two equations have the same structure and they are also not $t_0$ exact. Hence, they have to cancel among themselves and that can only happen if the following constraint is satisfied:
\begin{align}\label{d3.5}
\mathcal{H}^{b}{}^{\ell jmn}=-\varepsilon^{\ell p}\varepsilon^{jq}\varepsilon^{mr}\varepsilon^{ns}\mathcal{H}^{b}_{pqrs}\;.
\end{align}
The second and third expressions are $t_0$ exact and will cancel by taking $t_0$ operation on the following $J_{e^2\psi\bar{\psi}}$:
\begin{align}\label{d3.6}
J_{e^2\bar{\psi}\psi}&=-\frac{3}{32}e^a e^b \bar{\psi}^{j}\gamma_{ab}\gamma_{c}\psi_{k}\mathcal{G}^{c}_{j\ell}\varepsilon^{k\ell}\nonumber \\
&\;\;\; +\frac{1}{32}e^{a}e^{b}\bar{\psi}^{j}\gamma_{c}\gamma_{ab}\psi_{k}\mathcal{G}^{c}_{j\ell}\varepsilon^{k\ell} +\text{h.c}\;.
\end{align}
\subsection{Solving the remaining Bianchi and the final invariant density formula}

The remaining Bianchi that are to be solved are respectively $e^2 \psi^3$, $e^2\psi^2\bar{\psi}$, $e^3\psi^2$, $e^3\bar{\psi}{\psi}$, $e^4 {\psi}$ and their conjugates. However, it must be clear from the previous analyses that the Bianchi such as $e^2 \psi^3$, $e^3\psi^2$, $e^4 \psi$ and their conjugates will give rise to constraints and will not yield a new term in the four-form $J$. Where-as the Bianchi such as $e^2\psi^2\bar{\psi}$, its conjugate and $e^3\bar{\psi}{\psi}$  would give rise to constraints as well as yield new contributions to the four form $J$ which would be of the form $J_{e^3\psi}$ and its conjugate as well as $J_{e^4}$. However, the constraints that we would get from these Bianchi will not be independent ones. These constraints would be satisfied as a result of the earlier constraints (\ref{d2.6}, \ref{d3.5}) and closure of the supersymmetry algebra. This can be argued using a standard argument in superspace based on the Bianchi identity of the Bianchi identity (See appendix B.2 of \cite{Butter:2019edc}). We will revisit this argument in appendix-\ref{Bianchi}.

Upon solving the $e^2\bar{\psi}^2 \psi$ Bianchi, we would obtain the $J_{e^3\bar{\psi}}$ contribution to the four-form $J$ which is as follows:
\begin{align}\label{d4.1}
J_{e^3\bar{\psi}}&=\frac{1}{3}e^{a}e^{b}e^{c}\bar{\psi}_{i}\Upsilon^{d}_{j}\varepsilon^{ij}\varepsilon_{abcd}+\frac{1}{3}e^{a}e^{b}e^{c}\bar{\psi}_{i}\gamma^{d}\Psi_{j}\varepsilon^{ij}\varepsilon_{abcd}\;,
\end{align}
where,
\begin{align}\label{d4.2}
\Upsilon_{b}{}_{i}&=-\frac{1}{32}\gamma^{a}\sigma_{abi}\nonumber \\
\Psi_{i}&=- \frac{3}{64}\gamma^{a}\zeta_{ai}+\frac{1}{64}\gamma^{b}\beta_{bi}\;.
\end{align}
The components appearing above are related to the supersymmetry transformation of the components that appeared before as follows:
\begin{align}\label{d4.3}
\delta_{Q}^{L}\mathcal{B}_{abij}&=\bar{\epsilon}^{k}\nabla_{k}\mathcal{B}_{abij}=\bar{\epsilon}^{k}\lambda_{abijk}+\frac{2}{3}\varepsilon_{k(i}\bar{\epsilon}^{k}\sigma_{abj)}\nonumber \\
\delta_{Q}^{R}\mathcal{G}_{cij}&=\bar{\epsilon}_{k}\nabla^{k}\mathcal{G}_{cij}=\varepsilon^{kl}\bar{\epsilon}_{k}\xi_{cijl}-\frac{2}{3}\bar{\epsilon}_{(i}\zeta_{cj)} \nonumber \\
\delta_{Q}^{R}\bar{\mathcal{G}}_{cij}&=\bar{\epsilon}_{k}\nabla^{k}\bar{\mathcal{G}}_{cij}=\varepsilon^{kl}\bar{\epsilon}_{k}\alpha_{cijl}-\frac{2}{3}\bar{\epsilon}_{(i}\beta_{cj)}\;.
\end{align}
We have defined $\bar{\mathcal{G}}_{cij}$ as:
\begin{align}\label{d4.4}
\bar{\mathcal{G}}_{cij}&=\varepsilon_{ik}\varepsilon_{jl}{\mathcal{G}}_{c}^{kl}\;.
\end{align}
The conjugate $e^2\psi^2\bar{\psi}$ Bianchi will give rise to the conjugate $J_{e^3\psi}$. And finally the $e^3\psi\bar{\psi}$ Bianchi will give rise to the following:
\begin{align}\label{d4.5}
J_{e^4}&=-\frac{1}{24}e^{a}e^{b}e^{c}e^{d}\mathcal{F}\varepsilon_{abcd}+\text{h.c}\;,
\end{align}
where, $\mathcal{F}$ appears as the Lorentz invariant and SU(2) invariant component in the supersymmetry transformation of $\Psi_{i}$ as:
\begin{align}\label{d4.6}
\delta_{Q}^{L}\Psi_{j}&=-\frac{1}{2}\mathcal{Q}_{kj}\epsilon^{k}-\frac{1}{2}\varepsilon_{kj}\mathcal{F}\epsilon^{k}+\frac{1}{8}\gamma_{ab}\mathcal{P}^{ab}_{kj}\epsilon^{k}+\frac{1}{8}\gamma_{ab}\mathcal{R}^{ab}\epsilon^{k}\varepsilon_{kj}\;.
\end{align}
All the supercovariant terms will be contained in $J_{e^4}$. Using the above results, our final invariant action takes the following form:
\begin{align}\label{d4.7}
S=\int \mathcal{L} d^4 x \;,
\end{align}
where the Lagrangian density $\mathcal{L}$ is given by the density formula:
\begin{align}\label{d4.8}
e^{-1}\mathcal{L}&=-i\mathcal{F}+2i\bar{\psi}_{ai}\Upsilon^{a}_{j}\varepsilon^{ij}+2i\bar{\psi}_{ai}\gamma^{a}\Psi_{j}\varepsilon^{ij}+\frac{3i}{16}\bar{\psi}_{a}^{j}\gamma^{ab}\gamma^{c}\psi_{bk}\mathcal{G}_{cjl}\varepsilon^{kl}+\frac{i}{16}\bar{\psi}_{a}^{j}\gamma^{c}\gamma^{ab}\psi_{bk}\mathcal{G}_{cjl}\varepsilon^{kl}\nonumber \\
& \;\;\; -\frac{i}{4}\bar{\psi}_{a}^{i}\gamma^{ab}\psi_{b}^{j}\mathcal{D}^{kl}\varepsilon_{ik}\varepsilon_{jl}-\frac{i}{32}\bar{\psi}_{a}^{i}\gamma^{ab}\gamma^{cd}\psi_{b}^{j}\mathcal{B}_{cd}^{kl}\varepsilon_{ik}\varepsilon_{jl}-i\varepsilon^{abcd}\bar{\psi}_{ai}\gamma_{b}\psi_{c}^{j}\bar{\psi}_{dk}\Sigma_{jmn}\varepsilon^{km}\varepsilon^{in}+\text{h.c}\;.
\end{align}
Our basic building block in the above density formula is $\Sigma_{ijk}$ which satisfies the properties as explained below (\ref{density-12}) and also the constraints (\ref{d2.6}, \ref{d3.5}). The other components of the above density formula are found by taking consecutive supersymmetry transformations of $\Sigma_{ijk}$. The components transform among themselves under S-supersymmetry as shown below:
\begin{align}\label{d4.9}
\delta_S \Sigma_{ijk}&=0\nonumber \\
\delta_S \mathcal{D}_{ij}&=4\bar{\eta}^{k}\Sigma_{ijk} \nonumber\\
\delta_S \mathcal{B}_{ab}{}_{ij}&=8\bar{\eta}^{k}\gamma_{ab}\Sigma_{ijk} \nonumber\\
\delta_S \mathcal{G}_{a}{}_{ij}&=8\bar{\eta}_{k}\gamma_{a}\Sigma_{lij}\varepsilon^{kl} \nonumber\\
\delta_S \Upsilon_{d}{}_{j}&=\frac{1}{16}\gamma^{ab} \mathcal{B}_{ab}{}_{jk}\gamma_{d}\eta_{l}\varepsilon^{kl}-\frac{3}{8}\mathcal{G}_{d}{}_{jk}\eta^{k}+\frac{1}{8}\mathcal{G}^{g}{}_{jk}\gamma_{dg}\eta^{k} \nonumber\\
\delta_S \Psi_{j}&=-\frac{3}{4}\mathcal{D}_{jk}\eta_{l}\varepsilon^{kl}-\frac{3}{16}\gamma^{a}\mathcal{G}_{a}{}^{lm}\varepsilon_{jl}\varepsilon_{km}\eta^{k}+\frac{3}{16}\gamma^{a}\mathcal{G}_{a}{}_{jk}\eta^{k}\nonumber \\
\delta_S \mathcal{F}&=8\bar{\eta}_{i}\Psi_{j}\varepsilon^{ij}
\end{align}

We are interested in using the above density formula in obtaining invariant action of some known multiplets in $\mathcal{N}=2$ conformal supergravity. It seems that we can embed the 24+24 component real scalar multiplet of \cite{Hegde:2017sgl} in the above density formula to obtain an invariant action for the real scalar multiplet. It is also known from \cite{Hegde:2017sgl} that upon imposing suitable constraints on the real scalar multiplet one can embed the 8+8 component tensor multiplet in the 24+24 component real scalar multiplet. We will further use this embedding of tensor multiplet in the real scalar multiplet to get a new invariant action for the tensor multiplet coupled to $\mathcal{N}=2$ conformal supergravity.

In the next section, we will discuss the real scalar multiplet, its components and their supersymmetry transformation laws. We will also discuss the embedding of the tensor multiplet in the real scalar multiplet. In the subsequent section we will embed the real scalar multiplet into the density formula obtained in this section to get an invariant action for the real scalar multiplet. We will further use the embedding of the 8+8 component tensor multiplet to obtain a new higher derivative coupling of the tensor multiplet to $\mathcal{N}=2$ conformal supergravity.

\section{Real scalar multiplet}\label{Real}

The real scalar multiplet is a 24+24 component $\mathcal{N}=2$ matter multiplet that was originally found in flat space \cite{Howe:1982tm} in an attempt to understand off-shell formulation of $\mathcal{N}=2$ hypermultiplet. This multiplet was extended to $\mathcal{N}=2$ conformal supergravity in \cite{Hegde:2017sgl}. Further, in \cite{Hegde:2017sgl}, a consistent set of constraints was found to restrict the real scalar multiplet to an $8+8$ restricted real scalar multiplet. This restricted real scalar multiplet was shown to admit an embedding of the $8+8$ tensor multiplet. We will review the results from \cite{Hegde:2017sgl} briefly in this section.

Field content of the multiplet is given in Table-\ref{Table-Real-Scalar}. All the field components are invariant under special conformal transformations or K-transformation. Their $Q$ and $S$ transformation are given below\footnote{We have corrected some minor typos/omissions in the original paper \cite{Hegde:2017sgl}. The coefficient of $\bar{\Lambda}^{l}\Lambda^{m}\slashed{D}\Lambda_{(i}\varepsilon_{j|l|}\varepsilon_{k)m}$ in $\Gamma_{ijk}$ has been corrected and the term $-\frac{1}{16}\bar{\Lambda}^{p}\gamma^{ab}\Lambda^{q}\gamma_{ab}\Xi^{lmn}\varepsilon_{pq}\varepsilon_{il}\varepsilon_{jm}\varepsilon_{kn}$ which was missing in $\Gamma_{ijk}$has been added.}. 

\begin{table}[t]
	\caption{Field content of the $\mathcal{N}=2$ Real Scalar multiplet}\label{Table-Real-Scalar}
	\begin{center}
		\begin{tabular}{ | C{2cm}|C{2cm}|C{3cm}|C{2cm}|C{2cm}| }
			\hline
			Field & SU(2) Irreps & Restrictions &Weyl weight (w) & Chiral weight (c) \\ \hline
			$\phi$ & $\bf{1}$ & Real & 1 & 0 \\ \hline
			$E_{ij}$ & $\bf{3}$ & Complex &1 & -1  \\ \hline
			$S^a{}^i{}_j$ & $\bf{3}$ & $(S_a{}^{i}{}_{j})^{*}\equiv S_a{}_{i}{}^{j}=-S_a{}^{j}{}_{i}$ &1 & 0  \\ \hline
			$C_{ijkl}$ & $\bf{5}$ & $C^{ijkl}\equiv (C_{ijkl})^*=\varepsilon^{ip}\varepsilon^{jq}\varepsilon^{kr}\varepsilon^{ls}C_{pqrs}$ &2 & 0  \\ \hline
			$\Lambda_i$ & $\bf{2}$ & $\gamma_{5}\Lambda_i=\Lambda_i$&1/2 & 1/2  \\ \hline
			$\Xi_{ijk}$ & $\bf{4}$ & $\gamma_{5}\Xi_{ijk}=\Xi_{ijk}$ &3/2 &-1/2  \\ \hline
			
		\end{tabular}
	\end{center}
\end{table} 

\begin{align}\label{Susy-transf}
\delta \phi &= -\frac{\phi}{2}\bar{\epsilon}^{i}\Lambda_{i}+\text{h.c.}\;, \nonumber \\
\delta \Lambda^{i}&=-2\slashed{P}\epsilon^{i}-\left(\slashed{S}^{i}{}_{j}\epsilon^{j}+2\varepsilon^{ik}\varepsilon^{jl}\epsilon_{j}E_{lk}\right)-\frac{1}{2}\bar{\Lambda}^{i}\Lambda^{j}\epsilon_{j}-\frac{1}{4}\bar{\Lambda}^{j}\gamma_{a}\Lambda_{j}\gamma^{a}\epsilon^{i}+\frac{1}{8}\bar{\Lambda}^{i}\gamma_{ab}\Lambda^{j}\gamma^{ab}\epsilon_{j}\nonumber \\
& \;\;\;-2\eta^{i}\;,\nonumber\\
\delta S_{a}{}^{i}{}_{j}&=\bar{\epsilon}_{j}\gamma_{a}\chi^{i}+\frac{2}{3}\bar{\epsilon}_{j}\gamma_{a}\slashed{D}\Lambda^{i}-2\bar{\epsilon}_{j}D_{a}\Lambda^{i}-\frac{1}{3}\varepsilon^{li}\varepsilon^{nk}\bar{\epsilon}_{n}\gamma_{a}\Xi_{ljk}+\frac{1}{24}\bar{\epsilon}_{j}\gamma_{a}\gamma.T^{-}\Lambda_{k}\varepsilon^{ik} \nonumber\\
&\;\;\; -\frac{1}{3}\bar{\epsilon}_{j}\gamma_{a}\Lambda_{k}E_{lm}\varepsilon^{il}\varepsilon^{km}-\frac{2}{3}\bar{\epsilon}^{i}\gamma_{a}\slashed{S}^{k}{}_{j}\Lambda_{k}-\frac{1}{2}\bar{\epsilon}^{i}\slashed{S}^{k}{}_{j}\gamma_{a}\Lambda_{k}+\frac{1}{2}\bar{\epsilon}^{k}\gamma_{a}\slashed{S}^{i}{}_{j}\Lambda_{k} -\frac{2}{3}\bar{\epsilon}^{i}\gamma_{a}\slashed{P}\Lambda_{j}\nonumber \\
& \;\;\;-\bar{\epsilon}^{i}\slashed{P}\gamma_{a}\Lambda_{j}-\frac{1}{24}\bar{\Lambda}^{i}\Lambda^{k}\bar{\epsilon}_{j}\gamma_{a}\Lambda_{k}-\frac{1}{32}\bar{\Lambda}^{i}\gamma_{bc}\Lambda^{k}\bar{\epsilon}_{j}\gamma^{bc}\gamma_{a}\Lambda_{k}-\text{(h.c.;traceless)}\;,\nonumber\\
\delta E_{ij}&=2\bar{\epsilon}^{(l}\chi^{k)}\varepsilon_{ik}\varepsilon_{jl}-\frac{2}{3}\bar{\epsilon}^{(l}\slashed{D}\Lambda^{k)}\varepsilon_{ik}\varepsilon_{jl}+\frac{1}{3}\bar{\epsilon}^{k}\Xi_{ijk}-\frac{1}{12}\bar{\epsilon}^{k}\gamma.T^{-}\Lambda_{(i}\varepsilon_{j)k}+\frac{2}{3}\bar{\epsilon}^{k}\Lambda_{(i}E_{j)k}  \nonumber\\
& \;\;\;-2\bar{\epsilon}_{(i}\Lambda^{k}E_{j)k}-\frac{2}{3}\bar{\epsilon}^{k}\Lambda_{k}E_{ij}+\bar{\epsilon}_{k}\Lambda^{k}E_{ij}-\frac{1}{3}\bar{\epsilon}^{(l}\slashed{S}^{m)}{}_{k}\Lambda^{k}\varepsilon_{il}\varepsilon_{jm}-\frac{2}{3}\bar{\epsilon}^{(k}\slashed{P}\Lambda^{l)}\varepsilon_{ik}\varepsilon_{jl}\nonumber \\
& \;\;\;-\frac{1}{12}\bar{\epsilon}^{(l}\gamma_{a}\Lambda^{k)}\bar{\Lambda}^{m}\gamma^{a}\Lambda_{m}\varepsilon_{il}\varepsilon_{jk}\;, 
\nonumber \\
\delta \Xi_{ijk}&=\frac{3}{2}\epsilon_{mn}\epsilon_{lp}\left[D_{a}S^a{}^l{}_{(i}\delta^n_j\delta^p_{k)}-2\gamma^{ab}D_{a}S_{b}{}^l{}_{(i}\delta^n_j\delta^p_{k)}-\gamma.R(V)^l{}_{(i}\delta^n_j\delta^p_{k)} \right]\epsilon^m+{6}\slashed{D}E_{(ij}\epsilon_{k)} 
\nonumber \\
&\;\;\;-C_{ijkl}\epsilon^l-6E^{ln}E_{m(i}\varepsilon_{j|l|}\varepsilon_{k)n}\epsilon^{m} -6\slashed{P}E_{(ij}\epsilon_{k)}+3\slashed{S}^{m}{}_{(i}E_{jk)}\epsilon_{m}-6\slashed{S}^{m}{}_{(i}E_{j|m|}\epsilon_{k)} 
\nonumber\\
&\;\;\; +3S^{a}{}^{m}{}_{(i}S^{b}{}^{n}{}_{j}\varepsilon_{k)m}\varepsilon_{ln}\gamma_{ab}\epsilon^{l} + 3 P^{a}S_{a}{}^{l}{}_{(i}\varepsilon_{j|l|}\varepsilon_{k)m}\epsilon^{m}-\frac{3}{4}\gamma.T^{-}E^{lm}\epsilon^{n}\varepsilon_{(i|l|}\varepsilon_{j|m|}\varepsilon_{k)n}  
\nonumber\\
&\;\;\; +\bar{\Lambda}^{l}\gamma_{a}\slashed{D}\Lambda^{m}\gamma^{a}\epsilon_{(i}\varepsilon_{j|l|}\varepsilon_{k)m}+\frac{1}{4}\bar{\Lambda}^{l}\slashed{D}\Lambda_{(i}\epsilon^{m}\varepsilon_{j|l|}\varepsilon_{k)m}-{\frac{3}{8}}\bar{\Lambda}^{l}\slashed{D}\gamma_{ab}\Lambda_{(i}\gamma^{ab}\epsilon^{m}\varepsilon_{j|l|}\varepsilon_{k)m}
\nonumber \\
& \;\;\;-{\frac{1}{4}}\bar{\Lambda}_{(i}\slashed{D}\Lambda^{l}\epsilon^{m}\varepsilon_{j|l|}\varepsilon_{k)m} +{\frac{1}{8}}\bar{\Lambda}_{(i}\gamma_{ab}\slashed{D}\Lambda^{l}\gamma^{ab}\epsilon^{m}\varepsilon_{j|l|}\varepsilon_{k)m}-\frac{3}{2}\bar{\Lambda}_{(i}R(Q)_{ab}{}^{l}\gamma^{ab}\epsilon^{m}\varepsilon_{j|l|}\varepsilon_{k)m} 
\nonumber \\
&\;\;\;+\frac{1}{2}\bar{\Lambda}^{p}\Xi^{lmn}\epsilon^{q}\varepsilon_{il}\varepsilon_{jm}\varepsilon_{kn}\varepsilon_{pq}-\frac{1}{2}\bar{\Lambda}^{(m}\Xi^{np)l}\epsilon^{q}\varepsilon_{im}\varepsilon_{jn}\varepsilon_{kp}\varepsilon_{lq}-\frac{1}{8}\bar{\Lambda}_{l}\gamma_{ab}\Xi_{ijk}\gamma^{ab}\epsilon^{l}
\nonumber \\
&\;\;\;-\bar{\Lambda}_{(i}\Xi_{jk)l}\epsilon^{l}+\frac{1}{8}\bar{\Lambda}_{(i}\gamma_{ab}\Xi_{jk)l}\gamma^{ab}\epsilon^{l}-\frac{1}{2}\bar{\Lambda}^{l}\gamma_{a}\Xi_{ijk}\gamma^{a}\epsilon_{l}+\bar{\Lambda}^{l}\gamma_{a}\Xi_{l(ij}\gamma^{a}\epsilon_{k)}
\nonumber \\
&\;\;\; -\frac{3}{2}\bar{\Lambda}_{(i}\chi^{l}\epsilon^{m}\varepsilon_{j|l|}\varepsilon_{k)m}-\frac{3}{2}\bar{\Lambda}_{(i}\gamma_{ab}\chi^{l}\gamma^{ab}\epsilon^{m}\varepsilon_{j|l|}\varepsilon_{k)m}+\frac{3}{2}\bar{\Lambda}^{l}\chi_{(i}\epsilon^{m}\varepsilon_{j|l|}\varepsilon_{k)m}
\nonumber \\
& \;\;\; -3\bar{\Lambda}^{l}\gamma_{a}\chi^{m}\gamma^{a}\epsilon_{(i}\varepsilon_{j|l|}\varepsilon_{k)m}+\frac{1}{4}\varepsilon_{l(k}\bar{\Lambda}_{i}\Lambda_{j)}\gamma.T^{-}\epsilon^{l}-\frac{1}{8}\varepsilon_{l(k}\bar{\Lambda}^{l}\gamma_{a}\Lambda_{i}\gamma.T^{-}\gamma^{a}\epsilon_{j)} \nonumber \\
&\;\;\; +\frac{1}{2}\bar{\Lambda}^{(l}\Lambda^{m}E^{n)p}\epsilon^{q}\varepsilon_{il}\varepsilon_{jm}\varepsilon_{kn}\varepsilon_{pq}-\frac{1}{2}\bar{\Lambda}^{p}\Lambda^{(l}E^{mn)}\epsilon^{q}\varepsilon_{il}\varepsilon_{jm}\varepsilon_{kn}\varepsilon_{pq}+\frac{1}{2}\bar{\Lambda}_{l}\Lambda_{(i}E_{jk)}\epsilon^{l} 
\nonumber \\
&\;\;\; -\frac{1}{2}\bar{\Lambda}_{(i}\Lambda_{j}E_{k)l}\epsilon^{l}+\frac{1}{4}\bar{\Lambda}_{l}\gamma_{ab}\Lambda_{(i}E_{jk)}\gamma^{ab}\epsilon^{l} +\frac{1}{4}\bar{\Lambda}^{l}\gamma_{a}\Lambda_{l}\gamma^{a}\epsilon_{(i}E_{jk)}-\bar{\Lambda}^{l}\gamma_{a}\Lambda_{(i}\gamma^{a}\epsilon_{j}E_{k)l}
\nonumber \\
&\;\;\; -\frac{1}{2}\bar{\Lambda}^{l}\slashed{S}^{n}{}_{l}\Lambda_{(i}\epsilon^{m}\varepsilon_{j|n|}\varepsilon_{k)m}+\frac{1}{4}\varepsilon_{(i|n|}\varepsilon_{j|m|}\bar{\Lambda}^{l}\slashed{S}^{n}{}_{k)}\Lambda_{l}\epsilon^{m}-\frac{1}{8}\bar{\Lambda}^{l}\slashed{S}^{n}{}_{l}\gamma_{ab}\Lambda_{(i}\gamma^{ab}\epsilon^{m}\varepsilon_{j|n|}\varepsilon_{k)m}
\nonumber \\
&\;\;\;  +\frac{3}{16}\varepsilon_{(i|n|}\varepsilon_{j|m|}\bar{\Lambda}^{l}\slashed{S}^{n}{}_{k)}\gamma^{ab}\Lambda_{l}\gamma_{ab}\epsilon^{m}+\frac{1}{2}\bar{\Lambda}^{n}\Lambda^{m}\slashed{S}^{l}{}_{(i}\epsilon_{j}\varepsilon_{k)m}\varepsilon_{nl} +\frac{1}{2}\bar{\Lambda}^{l}\slashed{P}\Lambda_{(i}\epsilon^{m}\varepsilon_{j|l|}\varepsilon_{k)m}
\nonumber \\
&\;\;\; +\frac{1}{16}\bar{\Lambda}^{l}\gamma_{ab}\Lambda^{m}\slashed{S}^{n}{}_{(i}\gamma^{ab}\epsilon_{j}\varepsilon_{k)n}\varepsilon_{lm}+\frac{1}{2}\bar{\Lambda}^{l}\slashed{P}\gamma_{ab}\Lambda_{(i}\gamma^{ab}\epsilon^{m}\varepsilon_{j|l|}\varepsilon_{k)m}+\bar{\Lambda}^{l}\Lambda^{m}\slashed{P}\epsilon_{(i}\varepsilon_{j|l|}\varepsilon_{k)m}\nonumber \\
&\;\;\; +\frac{1}{8}\bar{\Lambda}^{m}\Lambda^{n}\bar{\Lambda}^{l}\gamma_{a}\Lambda_{l}\gamma^{a}\epsilon_{(i}\varepsilon_{j|m|}\varepsilon_{k)n} -\frac{1}{16}\bar{\Lambda}^{l}\Lambda^{m}\bar{\Lambda}_{n}\gamma_{ab}\Lambda_{(i}\gamma^{ab}\epsilon^{n}\varepsilon_{j|l|}\varepsilon_{k)m}\;,
\nonumber \\
\delta C_{ijkl}&=\bar{\epsilon}_{(i}\Gamma_{jkl)}+\varepsilon_{ip}\varepsilon_{jq}\varepsilon_{kr}\varepsilon_{ls}\bar{\epsilon}^{(p}\Gamma^{qrs)}+\left(\bar{\epsilon}_{m}\Lambda^{m}+\bar{\epsilon}^{m}\Lambda_{m}\right)C_{ijkl}\;.
\end{align}
where we have defined $P_a=\phi^{-1}D_a\phi$ and $\Gamma_{ijk}$ is given as follows.
\begin{align}\label{Gammadef}
\Gamma_{ijk}&=-2\slashed{D}\Xi_{ijk}-3D_{a}S^{a}{}^{n}{}_{(i}\Lambda^{m}\varepsilon_{j|n|}\varepsilon_{k)m}+6\slashed{D}E_{(ij}\Lambda_{k)}-2C_{ijkl}\Lambda^{l}+2\slashed{D}\Lambda_{(i}E_{jk)}
\nonumber \\
&\;\;\;  +2D_a\Lambda^l S^a{}^n{}_{(i}\varepsilon_{j|n|}\varepsilon_{k)l}-2\gamma^{ab}D_a\Lambda^l S_b{}^n{}_{(i}\varepsilon_{j|n|}\varepsilon_{k)l}-4\Xi^{lmn}E_{l(i}\varepsilon_{j|m|}\varepsilon_{k)n}+2\slashed{S}^{l}{}_{(i}\Xi_{jk)l}
\nonumber \\
&\;\;\;  +12 \chi_{(i}E_{jk)}-6\slashed{S}^{l}{}_{(i}\chi^{m}\varepsilon_{j|l|}\varepsilon_{k)m} -4\slashed{P}E_{(ij}\Lambda_{k)}-4P_aS^a{}^n{}_{(i}\varepsilon_{j|n|}\varepsilon_{k)l}\Lambda^l 
\nonumber \\
&\;\;\;  -2P_a\gamma^{ab}S_b{}^n{}_{(i}\varepsilon_{j|n|}\varepsilon_{k)l}\Lambda^l-2\slashed{S}^l{}_{(i}E_{jk)}\Lambda_l-2\slashed{S}^l{}_{(i}E_{j|l|}\Lambda_{k)}-4E_{(ij}\varepsilon_{k)m}\varepsilon_{ln}E^{mn}\Lambda^l 
\nonumber \\
&\;\;\; +12\varepsilon_{(i|m|}\varepsilon_{j|n|}E_{k)l}E^{mn}\Lambda^l+S_a{}^m{}_{(i}S^a{}^n{}_j\varepsilon_{k)m}\varepsilon_{ln}\Lambda^l+\gamma_{ab}S^{a m}{}_{(i}S^b{}^n{}_j\varepsilon_{k)m}\varepsilon_{ln}\Lambda^l 
\nonumber \\
&\;\;\;  +\frac{1}{2}\gamma\cdot T^+ E_{(ij}\varepsilon_{k)l}\Lambda^l+\frac{1}{4}\slashed{S}^l{}_{(i}\gamma\cdot T^-\Lambda_{j}\varepsilon_{k)l}+\frac{1}{4}\bar{\Lambda}^l\Lambda^m\slashed{D}\Lambda_{(i}\varepsilon_{j|l|}\varepsilon_{k)m}
\nonumber \\
&\;\;\;  +\frac{5}{4}\bar{\Lambda}^l\gamma^a\Lambda_{(i}\varepsilon_{j|l|}\varepsilon_{k)m}D_a\Lambda^m+\frac{5}{16}\bar{\Lambda}^l\gamma^{bc}\gamma^a\Lambda_{(i}\varepsilon_{j|l|}\varepsilon_{k)m}\gamma_{bc}D_a\Lambda^m
\nonumber \\
&\;\;\; +\frac{3}{2}\bar{\Lambda}^l\Lambda^m\chi_{(i}\varepsilon_{j|l|}\varepsilon_{k)m}-\frac{3}{2}\bar{\Lambda}^l\gamma^a\Lambda_{(i}\varepsilon_{j|l|}\varepsilon_{k)m}\gamma_a\chi^m +\frac{1}{2}\bar{\Lambda}^{q}\Lambda^{(p}\Xi^{ml)n}\varepsilon_{ip}\varepsilon_{jm}\varepsilon_{kl}\varepsilon_{qn}\nonumber \\
&\;\;\;-\frac{1}{16}\bar{\Lambda}^p\gamma^{ab}\Lambda^q\gamma_{ab}\Xi^{lmn}\varepsilon_{pq}\varepsilon_{il}\varepsilon_{jm}\varepsilon_{kn}+\frac{1}{2}\bar{\Lambda}^l\gamma_a\Lambda_{(i}\gamma^a\Xi_{jk)l}+\frac{3}{2}\bar{\Lambda}^l\Lambda^m\slashed{P}\Lambda_{(i}\varepsilon_{j|l|}\varepsilon_{k)m}\nonumber\\
&\;\;\; -\frac{1}{4}\bar{\Lambda}^l\Lambda^m \slashed{S}^n{}_{(i}\varepsilon_{j|n|}\varepsilon_{k)m}\Lambda_l-\frac{1}{32}\varepsilon_{mn}\bar{\Lambda}^m\gamma^{ab}\Lambda^n\gamma_{ab}\slashed{S}^l{}_{(i}\Lambda_j\varepsilon_{k)l}-\frac{1}{2}\bar{\Lambda}^l\Lambda^{(m}E^{pq)}\Lambda^n\varepsilon_{ip}\varepsilon_{jq}\varepsilon_{km}\varepsilon_{ln}
\nonumber \\
&\;\;\; -\bar{\Lambda}_{(i}\Lambda_j E_{k)l}\Lambda^l-\bar{\Lambda}_l\Lambda_{(i}E_{jk)}\Lambda^l+\frac{1}{8}\bar{\Lambda}^l\Lambda^m\bar{\Lambda}^n\gamma_a\Lambda_n\gamma^a\Lambda_{(i}\varepsilon_{j|l|}\varepsilon_{k)m}\;.
\end{align}
It is possible to reduce this multiplet to a restricted real scalar multiplet, by imposing a consistent set of constraints. In general the field $E_{ij}$ is a complex scalar field with chiral weight $-1$. Due to its non zero chiral weight, one can not demand it to satisfy a reality condition. However we can demand:
\begin{align}
E_{ij}=e^{-i\sigma/2}\mathcal{L}_{ij}
\end{align}
where $\sigma$ is a real scalar field with Weyl weight $0$ and $\mathcal{L}_{ij}$ is a triplet of scalars with Weyl weight $+1$ which satisfy the pseudo reality condition $\mathcal{L}^{ij}=\varepsilon^{ik}\varepsilon^{jl}\mathcal{L}_{kl}$. 

Let us define, $\bar{E}_{ij}=\varepsilon_{ik}\varepsilon_{jl}E^{kl}$. We can then put the above demand in the form of a constraint:
\begin{align}
\bar{E}_{ij}-e^{i\sigma}E_{ij}=0
\end{align}
This constraint reduces the six off-shell degrees of freedom present in $E_{ij}$ to four. Other constraints are obtained by supersymmetry transformation of the above constraint, and there exist $16+16$ consistent set of constraints which reduce the $24+24$ real scalar multiplet to an $8+8$ restricted real scalar multiplet. The precise form of the constraints will not be reviewed here. However, the gauge invariant objects of the $8+8$ tensor multiplet can be embedded in this $8+8$ real scalar multiplet. We present the precise embedding below. 

\begin{align}\label{iden}
\phi^{4}&=L^2\;, \nonumber\\
\Lambda^{i}&=-2L^{-2}L^{ij}\varphi_{j}\;, \nonumber\\
E_{ij}&=L^{-4}L_{ij}L^{kl}\bar{\varphi}_{k}\varphi_{l}-L^{-2}\bar{G}L_{ij}\;, \nonumber\\
S_{a}{}^{i}{}_{j}&=2L^{-2}H_{a}L^{ik}\varepsilon_{kj}+4L^{-4}L^{ik}L_{jm}\bar{\varphi}^{m}\gamma_{a}\varphi_{k}-L^{-2}\bar{\varphi}^{i}\gamma_{a}\varphi_{j}-\frac{1}{2}L^{-2}\delta^{i}_{j}\bar{\varphi}^{m}\gamma_{a}\varphi_{m}\;, \nonumber\\
&\;\;\; +L^{-2}\left(L^{ik}D_{a}L_{jk}-L_{jk}D_{a}L^{ik}\right)\;, \nonumber\\
\Xi_{ijk}&= -24L^{-6} L^{mn}\bar{\varphi}_m\varphi_nL_{(ij}L_{k)l}\varphi^l+6L^{-4}\bar{\varphi}_l\varphi_{(i}L_{jk)}\varphi^l-6L^{-4}L^{lm}\slashed{D}L_{l(i}L_{jk)}\varphi_m\nonumber\\
&\;\;\; +6L^{-4}L^{mn}\slashed{H}\varphi_nL_{(ij}\varepsilon_{k)m}+12L^{-4}\bar{G}L_{(ij}L_{k)l}\varphi^l+6L^{-2}\slashed{D}\varphi_{(i}L_{jk)}\nonumber\\
&\;\;\; +18L^{-2}L_{(ij}L_{k)l}\chi^l-\frac{3}{4}L^{-2}\gamma\cdot T^{-}\varphi^lL_{(ij}\varepsilon_{k)l}\;,\nonumber \\
C_{ijkl}&=-18L^{-2}DL_{(ij}L_{kl)}+6L^{-4}G\bar{G}L_{(ij}L_{kl)}+6L^{-4}H^aH_aL_{(ij}L_{kl)}\nonumber\\
&\;\;\;-12L^{-4}H^aD_aL^{mn}\varepsilon_{m(i}L_{jk}L_{l)n}-6L^{-2}D_aD^aL_{(ij}L_{kl)}+6L^{-4}L^{mn}D_aL_{mn}D^aL_{(ij}L_{kl)}\nonumber\\
&\;\;\;-3L^{-2}D^aL^{mn}D_aL_{mn}L_{(ij}L_{kl)}-9L^{-2}\bar{\chi}^m\varphi^n L_{(ij}\varepsilon_{k|m|}\varepsilon_{l)n}-36L^{-4}\bar{\chi}^m\varphi^nL_{(ij}L_{k|m|}L_{l)n}\nonumber\\
&\;\;\;-36L^{-4}L^{mn}\bar{\chi}_m\varphi_nL_{(ij}L_{kl)}+9L^{-2}\bar{\chi}_{(i}\varphi_jL_{kl)}+6L^{-4}G\bar{\varphi}_{(i}\varphi_jL_{kl)}\nonumber\\
&\;\;\;-18L^{-6}GL^{mn}\bar{\varphi}_m\varphi_nL_{(ij}L_{kl)} +6L^{-6}\bar{G}L_{mn}\bar{\varphi}^m\varphi^nL_{(ij}L_{kl)}-6L^{-4}\bar{G}\bar{\varphi}^m\varphi^nL_{(ij}\varepsilon_{k|m|}\varepsilon_{l)n}\nonumber\\
&\;\;\;-24L^{-6}\bar{G}\bar{\varphi}^m\varphi^nL_{(ij}L_{k|m|}L_{l)n}-6L^{-4}\bar{\varphi}^m\gamma^a\varphi_{(i}L_{jk}D_aL_{l)m}\nonumber\\
&\;\;\;
+36L^{-6}\bar{\varphi}^m\gamma^a\varphi_nL^{ns}D_aL_{s(i}L_{jk}L_{l)m}-6L^{-4}\bar{\varphi}^m\gamma^a\varphi_mD_aL_{(ij}L_{kl)}\nonumber\\
&\;\;\;+6L^{-4}\bar{\varphi}^m\slashed{H}\varphi_{(i}L_{jk}\varepsilon_{l)m}-36L^{-6}\bar{\varphi}^m\slashed{H}\varphi_nL^{ns}L_{(ij}L_{k|m|}\varepsilon_{l)s}\nonumber\\
&\;\;\;-12L^{-4}L^{st}\varphi_s\slashed{D}\varphi^mL_{(ij}\varepsilon_{k|t|}\varepsilon_{l)m}-12L^{-4}\bar{\varphi}^m\slashed{D}\varphi_{(i}L_{jk}L_{l)m}\nonumber\\
&\;\;\;+\frac{3}{4}L^{-4}\varepsilon_{mn}\bar{\varphi}^m\gamma\cdot T^-\varphi^nL_{(ij}L_{kl)}+\frac{3}{4}L^{-4}\varepsilon^{mn}\bar{\varphi}_m\gamma\cdot T^+\varphi_nL_{(ij}L_{kl)}\nonumber\\
&\;\;\;-36L^{-6}\bar{\varphi}^m\varphi^n\bar{\varphi}_m\varphi_nL_{(ij}L_{kl)}-36L^{-6}L_{mn}\bar{\varphi}^m\varphi^n\bar{\varphi}_{(i}\varphi_jL_{kl)}+36L^{-6}\bar{\varphi}^m\varphi^n\bar{\varphi}_m\varphi_{(i}L_{jk}L_{l)n}\nonumber\\
&\;\;\;+90L^{-8}L^{mn}\bar{\varphi}_m\varphi_nL_{st}\bar{\varphi}^s\varphi^tL_{(ij}L_{kl)}\;,
\end{align}
 
When the real scalar multiplet fields are expressed in terms of the $8+8$ tensor multiplet fields as given above, one can verify that they obey the $16+16$ set of constraints given in \cite{Hegde:2017sgl}. We will show in the next section that the $24+24$ real scalar multiplet can be embedded into the density formula presented in the previous section, thereby allowing us to construct a new action for the tensor multiplet in $\mathcal{N}=2$ conformal supergravity.
\section{Invariant action for the real scalar multiplet}\label{Real_action}
In order to apply the density formula obtained in section-\ref{density} for obtaining the invariant action for the real scalar multiplet, we need to find a suitable combination of the fields belonging to the real scalar multiplet that has all the properties satisfied by $\Sigma_{ijk}$ which is the block that appears in the density formula with the maximum number of gravitino (\ref{density-12}). It turns out that the following choice can be made:
\begin{align}\label{R1}
\Sigma_{ijk}=i\Gamma_{ijk}+4iC_{ijk\ell}\Lambda^\ell\;,
\end{align}
where, $\Gamma_{ijk}$ is defined in (\ref{Gammadef}). Upon taking the supersymmetry transformations of $\Sigma_{ijk}$ up to terms that are quadratic in fermions, we obtain the following expressions for $\mathcal{A}^{ab}_{ijkl}$ and $\mathcal{H}^{a}_{ijkl}$:
\begin{align}\label{R2}
\mathcal{H}^a_{ijkl}&=-4iD^aC_{ijkl}+8iP^aC_{ijkl}+8iS^{am}{}_{(i}C_{jkl)m}-2i\bar{\Lambda}_{(i}\gamma^a\slashed{D}\Xi_{jkl)}\nonumber\\
&\;\;\; +2i\bar{\Lambda}^p\gamma^a\slashed{D}\Xi^{stu}\varepsilon_{s(i}\varepsilon_{j|t|}\varepsilon_{k|u|}\varepsilon_{l)p}\nonumber\\
\mathcal{A}^{ab}_{ijkl}&=0\;.
\end{align}
Without explicitly evaluating the remaining terms, we can argue that they indeed obey the constraints (\ref{d2.6}, \ref{d3.5}). Firstly, it is clear that the above expressions obey the constraints. It turns out that they are also S-invariant. This follows from the closure of the supersymmetry algebra, in particular $[\delta_Q,\delta_S]$ on $\Sigma_{ijk}$.

Secondly, any term that are cubic in fermions in $\mathcal{H}^a_{ijkl}$ and $\mathcal{A}^{ab}_{ijkl}$ has to necessarily contain a bare $\Lambda^{i}$ or $\Lambda_{i}$ which transforms under S-supersymmetry to a bare S-supersymmetry parameter (\ref{Susy-transf}). Hence, such a term would be related by S-supersymmetry to terms that are quadratic in fermions and would automatically satisfy the constraints because the terms that are quadratic in fermions already satisfy the constraints. Similarly terms that are quartic in fermions would be related to the terms cubic in fermions by S-supersymmetry and hence they would also satisfy the constraints. As a result, we can argue that the constraints are completely satisfied without actually evaluating all the terms.

Alternatively one can also see that the constraints are satisfied as a result of the closure of supersymmetry algebra on the $C_{ijkl}$ component of the real scalar multiplet. The closure also gives some more constraints apart from the constraints required by the density formula. For example, we find that the embedding of the real scalar multiplet into the abstract multiplet constituting the density formula would also satisfy the following constraint:
\begin{align}\label{R3}
\bar{{G}}^{a}_{ij}\equiv \varepsilon_{ik}\varepsilon_{jl}{G}^{a}{}^{kl}=-{G}^{a}_{ij}\;,
\end{align}
where, $G^{a}_{ij}$ is the S-invariant combination:
\begin{align}\label{R3.1}
{G}^{a}_{ij}\equiv \mathcal{G}^{a}_{ij}+4\bar{\Lambda}_{k}\gamma^{a}\Sigma_{lij}\varepsilon^{kl}\;.
\end{align}
This is expected since the real scalar multiplet is smaller than the abstract multiplet constituting the density formula. However, it serves as a consistency check on our calculations.

We are interested in giving the final action that contains only bosonic terms. Hence, we will adopt the following minimalistic route. From $\Sigma_{ijk}$ given in (\ref{R1}), we will obtain $\mathcal{G}^{a}_{ij}$ up to terms quadratic in fermions using (\ref{d3.2}). Following this, we will obtain $\Psi_{i}$ up to terms linear in fermions using (\ref{d4.2}) and (\ref{d4.3}). And subsequently we obtain the bosonic $\mathcal{F}$ using (\ref{d4.6}), which will give us the Lagrangian density $\mathcal{L}$ (\ref{d4.8}). In order to avoid cluttering, we give the final result for bosonic $\mathcal{L}$. We perform some integration by parts thereby generating some bare K-gauge field. We will present all the intermediate results in appendix-\ref{inter}.
\begin{align}\label{R4}
e^{-1}\mathcal{L}&=-\frac{160}{9}D_{a}E^{ij}D_{a}E_{ij}+\frac{4}{3}D^a T^{-}_{ac}D^b T^{+}_{bc}-\frac{7}{9}D_a S^{a}{}^{i}{}_{j}D_{b}S^{b}{}^{j}{}_{i}+3D_a S_{b}{}^{i}{}_{j}D^{b}S^{a}{}^{j}{}_{i}\nonumber\\
&\;\;\; +24 D^2+\frac{4}{9}C_{ijkl}C^{ijkl}-4R(V)_{ab}{}^{i}{}_{j}R(V)^{ab}{}^{j}{}_{i}+{\frac{80}{3}}E_{ij}D_{a}E^{jk}S^{a}{}^{i}{}_{k}\nonumber\\
&\;\;\; +{\frac{67}{24}}S^{a}{}^{i}{}_{j}S^{b}{}^{j}{}_{k}D_{a}S_{b}{}^{k}{}_{i}+{\frac{41}{12}}D_{a}E^{ij}S_{b}{}_{ij}T^{-}{}^{ab}-\frac{5}{4}E^{ij}D_{a}S_{b}{}_{ij}T^{-}{}^{ab}\nonumber\\
& \;\;\; +{\frac{56}{9}}P^b S_{b}{}^{i}{}_{j}D^a S_{a}{}^{j}{}_{i}-{\frac{8}{3}} P_b S_{a}{}^{i}{}_{j}D^a S^{b}{}^{j}{}_{i}+\frac{320}{9}P^a E_{ij}D_{a}E^{ij}\nonumber\\
& \;\;\; +{\frac{8}{3}}P^aT^{+}_{ac}D_{b}T^{-}{}^{bc}+{5}S_{a}{}^{i}{}_{j}S_{b}{}^{j}{}_{k}R(V)^{ab}{}^{k}{}_{i}+\frac{5}{3} R(V)_{ij}\cdot T^{-}E^{ij}\nonumber\\
& \;\;\; -{\frac{1}{3}}S_{a}{}^{ij}S^{a}{}^{kl}C_{ijkl}+\frac{16}{3}C^{ijkl}E_{ij}\bar{E}_{kl}-{\frac{13}{2}}D S_{a}{}^{i}{}_{j}S^{a}{}^{j}{}_{i}-{3}f_{a}{}^{a}S_{b}{}^{i}{}_{j}S^{b}{}^{j}{}_{i}\nonumber\\
&\;\;\; -{\frac{10}{3}}f^{ab}S_{a}{}^{i}{}_{j}S_{b}{}^{j}{}_{i} +{\frac{56}{9}}P^a P^b S_{a}{}^{i}{}_{j}S_{b}{}^{j}{}_{i}+{\frac{4}{3}}P^a P_{a} S_{b}{}^{i}{}_{j}S^{b}{}^{j}{}_{i}-\frac{160}{9}P^a P_{a}E_{ij}E^{ij}\nonumber\\
&\;\;\; +{\frac{4}{3}}P_{a}P_{b}T^{-}{}^{ac}T^{+}{}^{b}{}_{c} +\frac{20}{9}P^a S_{a}{}^{i}{}_{j}E_{ik}E^{jk}-\frac{41}{12}P_a S_{b}{}_{ij}E^{ij}T^{-}{}^{ab}\nonumber \\
&\;\;\; +\frac{188}{9}E^{mn}E_{mn}E^{kl}E_{kl}-\frac{184}{9}E^{mn}E_{mk}E^{kl}E_{ln}+\frac{2}{3}T^+\cdot T^+ \bar{E}^{mn}E_{mn}\nonumber\\
&\;\;\; +\frac{17}{6}S_{amn}S^{akl}E_{kl}E^{mn}+\frac{13}{6}E^{mn}E_{mn}S^{akl}S_{akl}+\frac{2}{3}S^{akl}S_{amn}S^{bmn}S_{bkl}\nonumber\\
&\;\;\; -\frac{5}{6}S^{amn}S_{amn}S^{bkl}S_{bkl}+\frac{55}{6}T^{-ab}S_{alm}S^{bl}{}_nE^{mn}-\frac{47}{36}S^{amn}S^b_{mn}T^-_{ac}T^+{}_b{}^c+\text{h.c}
\end{align}
In the above equation, we have used the following definition:
\begin{align}\label{R5}
&\bar{E}_{ij}\equiv \varepsilon_{ik}\varepsilon_{jl}E^{kl}\;, \quad \bar{E}^{ij}\equiv \varepsilon^{ik}\varepsilon^{jl}E_{kl}\nonumber \\
&S_{a}{}_{ij}\equiv S_{a}{}^{k}{}_{i}\varepsilon_{kj}\;, \quad S_{a}^{ij}\equiv \varepsilon^{ik}\varepsilon^{jl}S_{a}{}_{kl}\nonumber \\
&R(V)_{ab}{}_{ij}=R(V)_{ab}{}^{k}{}_{i}\varepsilon_{kj}
\end{align}

\section{New higher derivative coupling of tensor multiplet in $\mathcal{N}=2$ conformal supergravity}\label{tensor_action}
In this section we will use the embedding of the tensor multiplet in real scalar multiplet (\ref{iden}) and the invariant action for the real scalar multiplet (\ref{R4}) to obtain new higher derivative action for the tensor multiplet in $\mathcal{N}=2$ conformal supergravity. The bosonic terms in the action are:
\begin{align}\label{tensor}
e^{-1}\mathcal{L}&=\frac{1}{4}L^{-2}T^+\cdot T^+ \bar{G}^2-\frac{4}{3}L^{-2}T^-_{ac}T^+{}_b{}^cH^aH^b-16 L^{-2}DG\bar{G}-\frac{16}{3}L^{-2}G\bar{G}R\nonumber\\
&\;\;\; -2L^{-2}(H^aH_a)D-\frac{5}{3}L^{-2}R(H^aH_a)^2+\frac{5}{2}L^{-2}H_aH_bR^{ab}-\frac{5}{8}L^{-2}\bar{G}L^{ij}R(V)_{ij}\cdot T^-\nonumber\\
&\;\;\;-\frac{41}{16}L^{-2}D_a\bar{G}H_bT^{-ab}+\frac{15}{16}L^{-2}\bar{G}D_aH_bT^{-ab}+45D^2+\frac{3}{2}R(V)_{ab}{}^{ij}R(V)^{ab}{}_{ij} \nonumber\\
&\;\;\;+\frac{1}{2}D^a T^{-}_{ac}D^b T^{+}_{bc}-\frac{20}{3}L^{-2}D_aGD^a\bar{G}-\frac{9}{2}L^{-2}D_aH_bD^bH^a-\frac{41}{16}L^{-2}D_aGH_bT^{-ab}\nonumber\\
&\;\;\;+\frac{15}{16}L^{-2}GD_aH_bT^{-ab}+16 L^{-4}(G\bar{G})^2+L^{-4}(H^aH_a)^2+\frac{43}{2}L^{-4}(H^aH_a)(G\bar{G})\nonumber\\
&\;\;\;+\frac{1}{2}L^{-1}D^aLT^+_{ac}D_bT^{-ac}+11 L^{-3}D_aH_bH^aD^bL+32L^{-3}H^aD_bH_aD^bL\nonumber\\
&\;\;\;+52 L^{-3}D^aL(D_aG)\bar{G}+\frac{93}{32}L^{-3}GH_bD_aLT^{-ab}+9L^{-4}\varepsilon_{ik}D^bH^aD_aL^{ij}D_bL^{kl}L_{jl}\nonumber\\
&\;\;\; -\frac{317}{32}L^{-2}H_aD_bL^{ij}R(V)^{ab}{}_{ij}+\frac{317}{32}L^{-3}H_aD_bLL^{ij}R(V)^{ab}{}_{ij}+16L^{-2}D^aL^{ij}D_aL_{ij}D\nonumber\\
&\;\;\; - 26L^{-2}D_aLD^aLD+24L^{-2}D^aL^{ij}L_{ij}D-\frac{15}{4}L^{-2}D_aL^{im}D_bL_{km}R(V)_{ab}{}^k{}_i\nonumber\\
&\;\;\;+\frac{5}{4}L^{-2}R^{ab}D_aL_{ij}D_bL^{ij}-\frac{5}{4}L^{-2}R^{ab}D_aLD_bL+\frac{1}{6}L^{-2}RD_aL_{ij}D^aL^{ij}\nonumber \\
&\;\;\; -\frac{1}{6}L^{-2}RD^aLD_aL-\frac{83}{8}L^{-4}G\varepsilon_{ik}D_aL^{ij}D_bL^{kl}L_{jl}T^{-ab}-\frac{47}{48}L^{-4}D_aL_{ij}D_bL^{ij}T^{-ac}T^{+b}{}_c\nonumber\\
&\;\;\;+\frac{53}{48}L^{-4}L^{-4}H^aH^bD_aLD_bLT^{-ac}T^{+b}{}_c-\frac{67}{16}L^{-4}\varepsilon_{ik}D^aL^{ij}D_bL^{kl}L_{jl}D_aH_b\nonumber\\
&\;\;\;-\frac{37}{96}L^{-4}H^aH^bD_aLD_bL- 23L^{-4}L^{kl}H^aH_aD_bLD^bL-\frac{667}{192}L^{-4}H^aH^bD_aL_{ij}D_bL^{ij}\nonumber\\
&\;\;\;-\frac{13}{2}L^{-4}H^aH_aD_bL_{ij}D^bL^{ij}-6L^{-4}H^aH_aD^2L_{ij}L^{ij}-\frac{1643}{24}L^{-4}G\bar{G}D_aLD^aL\nonumber\\
&\;\;\;+\frac{323}{24}L^{-4}G\bar{G}D_aL_{mn}D^aL^{mn}-16L^{-4}G\bar{G}L_{ij}D^2L^{ij}\nonumber\\
&\;\;\; +\frac{313}{48}L^{-5}H^a (D^{b}L)\varepsilon_{il}L^{ij}D_{a}L_{jk}D_{b}L^{kl}-\frac{7}{3}L^{-4}H^{a}\varepsilon_{il}L^{ij}D_{a}L_{jk}D^2 L^{kl}\nonumber\\
&\;\;\;+\frac{43}{12}L^{-2}D^2L^{ij}D^2L_{ij}+\frac{9}{4}L^{-4}D_aD_bL_{ij}D^bD^aL^{ij}-\frac{9}{4}L^{-4}D_aD_bL^{ij}L_{ij}D^bD^aL^{kl}L_{kl}\nonumber\\
&\;\;\;+\frac{29}{12}L^{-4}D^2L^{ij}L_{ij}D^2L^{kl}L_{kl}-\frac{32}{3}L^{-3}L^{-3}D^aLD_aL^{ij}D^2L_{ij}+\frac{13}{6}L^{-5}D^aLD_aLL^{ij}D^2L_{ij}\nonumber\\
&\;\;\;+\frac{37}{8}D^2L_{ij}L^{ij}D_aL_{kl}D^aL^{kl}+\frac{26}{3}L^{-4}D_aLD_bLD^aD^bL_{ij}L^{ij}\nonumber\\
&\;\;\; -\frac{211}{32}L^{-4}D_aD_bL^{ij}L_{ij}D^aL^{kl}D^bL_{kl}+9L^{-4}D_aD_bL^{ij}D^aL_{ik}L_{jl}D^bL_{kl}\nonumber\\
&\;\;\; -\frac{67}{16}L^{-4}L^{ij}D^aL_{ik}D^bL_{jl}D_aD_bL^{kl}+\frac{233}{96}L^{-3}D_aL_{ij}D_bLD^aD^bLD^aD^bL^{ij}\nonumber \\
&\;\;\; +\frac{67}{32}L^{-3}D_bL_{ij}D_aLD^aD^bL^{ij}+\frac{43}{24}L^{-4}D_aLD^aLD_bL_{ij}D^bL^{ij}\nonumber\\
&\;\;\;+\frac{793}{96}L^{-4}D_aLD_bLD^aL^{ij}D^bL_{ij}-\frac{37}{3}L^{-4}D^aLD_aLD^bLD_bL\nonumber\\
&\;\;\;-\frac{5}{32}L^{-4}D_aL_{ij}D^aL^{ij}D_bL_{kl}D^bL^{kl}-\frac{35}{32}L^{-4}D^aL^{ij}D^bL_{ij}D_aL^{kl}D_bL_{kl}+\text{h.c}\;.
\end{align}
To bring the action in this form we replaced the dependent K-gauge field $f_{a}{}^{b}$ as shown below:
\begin{align}\label{t1}
f_{a}{}^{b}&=\frac{1}{2}R_{a}{}^{b}-\frac{1}{4}(D+\frac{1}{3}R)\delta_{a}{}^{b}-\frac{i}{2}\tilde{R}(A)_{a}{}^{b}+\frac{1}{8}T^{-}_{ac}T^{+}{}^{bc}\;,
\end{align}
where,
\begin{align}\label{t2}
R_{a}{}^{b}&\equiv (R(M)|_{f=0})_{ac}{}^{bc}\;, \nonumber \\
R&\equiv R_{a}{}^{a}\;.
\end{align}
The first of the above equations means that we first set $f=0$ in the expression for the supercovariant curvature $R(M)_{ab}{}^{cd}$ and then take the contraction between the Lorentz indices. If we confine ourselves to the bosonic background and upon using the gauge fixing condition for special conformal transformation ($b_\mu=0$), the $R_{a}{}^{b}$ and $R$ becomes the standard Ricci tensor and Ricci scalar respectively.
\section{Conclusions and Future Directions}
Matter multiplets play an important role in the superconformal approach for constructing off-shell supergravity theories. They are commonly used as compensators in gauge fixing the additional symmetries present in conformal supergravity to get the physical Poincar{\'e} supergravity. They can also be added as extra matter multiplets in the supergravity theories. Hence the study of various matter multiplets and their coupling to conformal supergravity plays an important role in the construction of higher derivative invariants in supergravity theories using the superconformal approach. The matter multiplets that play an important role in the construction of $\mathcal{N}=2$ supergravity theories are vector multiplet, tensor multiplet, non linear multiplet or hypermultiplet. For the construction of off-shell $\mathcal{N}=2$ supergravity theories typically vector multiplet and either non-linear or tensor multiplets are used as compensators. The coupling of tensor multiplet to conformal supergravity has been discussed earlier in the literature either using the linear-vector density formula or chiral density formula \cite{deWit:1982na, deWit:2006gn}. In this paper, we have discussed an alternate density formula based on a fermionic multiplet whose lowest component is dimension-5/2 spinor transforming in the $\bf{4}$ irrep of the underlying SU(2)-R symmetry and is a superconformal primary field. We arrived at this density formula using the covariant superform approach which was discussed in details in \cite{Butter:2019edc} for constructing an invariant action for $\mathcal{N}=4$ conformal supergravity. Using the new density formula, we followed a series of steps to construct a new higher derivative coupling of the tensor multiplet to conformal supergravity. Analogous new linear multiplet actions higher order in derivatives was found in six dimensions \cite{Butter:2018wss}. It seems plausible that the tensor multiplet action (\ref{tensor}) that we have obtained may be related to the six dimensional result by a dimensional reduction, although it might be a non trivial exercise to establish this connection. Tensor multiplets can be used as a compensating multiplet to go from conformal supergravity to Poincar{\'e} supergravity and hence the results of this paper will induce new higher derivative corrections to Poincar{\'e} supergravity that were not present in the earlier literature. Higher derivative corrections to supergravity play an important role in the discussion of black hole entropy and its matching with the microscopic results originating from string theory. It would be interesting to see what are the implications of the results of this paper on the study of black hole entropy. Study of tensor multiplets as a matter multiplet instead of compensating multiplet is also important as has been discussed in \cite{Cribiori:2018xdy}. It would be interesting to see the effect of the new higher derivative couplings on the results of \cite{Cribiori:2018xdy}. We would also like to see if the applicability of the new density formula can be taken beyond real scalar multiplet and tensor multiplet to find new higher derivative actions for other multiplets, for e.g vector multiplet or non linear multiplet.
\acknowledgments

We thank Gabriele Tartaglino Mazzucchelli for useful correspondence. SH thanks Amitabh Virmani and Kedar Kolekar for hospitality at the Chennai Mathematical Institute, and Nemani Suryanarayana for hospitality at the Institute of Mathematical Sciences, Chennai during the course of this work. This work is supported by SERB grant CRG/2018/002373, Government of India. 
\appendix
\section{$\mathcal{N}=2$ conformal supegravity in four dimensions}\label{N2Weyl}
In $\mathcal{N}=2$ conformal supergravity in four dimensions, the standard Weyl multiplet has 24+24 off-shell degrees of freedom with the field content described in Table-\ref{Table-Weyl}. The fields $V_\mu{}^i{}_j$ and $A_\mu$ are the gauge fields corresponding to the $SU(2)_R\times U(1)_R$ $R$-symmetry which is a necessary feature of the $SU(2,2|2)$ superconformal algebra. Fields form irreps under this $SU(2)_R$ and we will follow the chiral convention where raising or lowering the $SU(2)_R$ indices is achieved by complex conjugation and reverses the chirality of the fermions as well as the chiral weight of the corresponding fields. Weyl weight and chiral weight denote the weights with which the fields transform under dilatation and $U(1)_R$ symmetries respectively. 

\begin{table}[t]
	\caption{Field content of the $\mathcal{N}=2$ Weyl multiplet}\label{Table-Weyl}
	\begin{center}
		\begin{tabular}{ | C{2cm}|C{2cm}|C{3cm}|C{2cm}|C{2cm}| }
			\hline
			Field & SU(2) Irreps & Restrictions &Weyl weight (w) & Chiral weight (c) \\ \hline
			$e_{\mu}{}^{a}$ & $\bf{1}$ & Vielbein & -1 & 0 \\ \hline
			$V_{\mu}{}^{i}{}_{j}$ & $\bf{3}$ & $(V_{\mu}{}^{i}{}_{j})^{*}\equiv V_{\mu}{}_{i}{}^{j}=-V_{\mu}{}^{j}{}_{i}$ SU(2)$_R$ gauge field &0 & 0  \\ \hline
			$A_{\mu}$ & $\bf{1}$ & U(1)$_R$ gauge field &0 & 0  \\ \hline
			$b_{\mu}$ & $\bf{1}$ & dilatation gauge field &0 & 0  \\ \hline
			$T^{-}_{ab}$ & $\bf{1}$ & Anti self-dual i.e $T^{-}_{ab}=-\frac{1}{2}\varepsilon_{abcd}T^{-}{}^{cd}$ &1 & -1  \\ \hline
			$D$ & $\bf{1}$ & Real &2 & 0  \\ \hline
			$\psi_{\mu}{}^{i}$ & $\bf{2}$ & $\gamma_{5}\psi_{\mu}{}^{i}=\psi_{\mu}{}^{i}$&-1/2 & -1/2  \\ \hline
			$\chi^{i}$ & $\bf{2}$ & $\gamma_{5}\chi^{i}=\chi^{i}$ &3/2 &-1/2  \\ \hline

		\end{tabular}
	\end{center}
\end{table}

The superconformal algebra contains two types of supersymmetries, $Q$ and $S$ supersymmetries. The field $\psi_\mu^i$ is the gauge field corresponding to $Q$-supersymmetry. The field $\phi_\mu^i$ is the gauge field for $S$ supersymmetry and it is a dependent gauge field due to constraints imposed on superconformal curvature to obtain a minimal representation of the superconformal algebra. The gauge fields for local rotations and conformal boosts, $\omega_\mu^{ab}$ and $f_\mu^a$ also occur as dependent gauge fields. In $\mathcal{N}=2$ conformal supergravity in four dimensions, the conventional set of constraints are given by,

\begin{align}\label{W}
R_{\mu\nu}(P)^a&=0\;, \nonumber\\
\gamma^\mu R_{\mu\nu}(Q)^i+\frac{3}{2}\gamma_\nu\chi^i&=0\;,\nonumber\\
e^\nu_bR_{\mu\nu}(M)_a{}^b-i\tilde{R}_{\mu a}(A)+\frac{1}{4}T^+_{ab}T^-_\mu{}^b-\frac{3}{2}De_{\mu a}&=0\;,
\end{align} 
where $R(P)$, $R(M)$, $R(A)$ and $R(Q)$ are curvatures corresponding to local translation, local rotation, $U(1)$ $R$-symmetry and $Q$ supersymmetry respectively.

With the above constraints, the following soft algebra is satisfied on fields:
\begin{align}\label{algebra}
\left[\delta_Q(\epsilon_1),\delta_Q(\epsilon_2)\right]&=\delta^{(cov)}(\xi)+\delta_{M}(\varepsilon)+\delta_{K}(\Lambda_K)+\delta_S(\eta)+\delta_{gauge}\;,
\nonumber \\
&\nonumber \\
\left[\delta_S(\eta),\delta_Q(\epsilon)\right]&=\delta_{M}(\bar{\eta}_{i}\gamma^{ab}\epsilon^{i}+\text{h.c})+\delta_{D}(\bar{\eta}^{i}\epsilon_{i}+\text{h.c})+\delta_A(i\bar{\eta}_{i}\epsilon^{i}+\text{h.c})
\nonumber \\
&\quad +\delta_{V}(-2\bar{\eta}^{i}\epsilon_{j}-(\text{h.c ; traceless}))\;, \nonumber \\
\left[\delta_S(\eta_1),\delta_S(\eta_2)\right]&=\delta_{K}(\bar{\eta}_{2i}\gamma^a \eta_{1}^{i}+\text{h.c})\;.
\end{align}
where $\delta_{\xi}$, $\delta_{M}$, $\delta_{D}$, $\delta_A$ , $\delta_{V}$, $\delta_{K}$, $\delta_{S}$ are covariant general coordinate transofrmation, local rotation, dilatation, $U(1)_R$, $SU(2)_R$, special conformal boosts and $S$-supersymmetry transformations with the following field dependent transformation parameters. 
\begin{align}\label{parameters_algebra}
\xi^{\mu}&=2\bar{\epsilon}_{2}^{i}\gamma^{\mu}\epsilon_{1i}+\text{h.c.}\;,
\nonumber \\
\varepsilon_{ab}&=\varepsilon^{ij}\bar{\epsilon}_{1i}\epsilon_{2j}T^{-}_{ab}+\text{h.c.}\;,\nonumber \\
\Lambda_{K}^{a}&=\varepsilon^{ij}\bar{\epsilon}_{1i}\epsilon_{2j}D_{b}T^{-}{}^{ba}-\frac{3}{2}\bar{\epsilon}_{2}^{i}\gamma^{a}\epsilon_{1i}D+\text{h.c.}\;,
\nonumber \\
\eta^{i}&=6\bar{\epsilon}_{1}^{[i}\epsilon_{2}^{j]}\chi_{j}\;.
\end{align}
Matter multiplets coupled to $\mathcal{N}=2$ conformal supergravity satisfy the above algebra. Off-shell multiplets such as vector multiplet, tensor multiplet, nonlinear multiplet or real scalar multiplet satisfy the above algebra without the need to impose equations of motion. 

\begin{table}[t]
	\caption{Field content of the $\mathcal{N}=2$ tensor multiplet}\label{Table-Tensor}
	\begin{center}
		\begin{tabular}{ | C{2cm}|C{2cm}|C{3cm}|C{2cm}|C{2cm}| }
			\hline
			Field & SU(2) Irreps & Restrictions &Weyl weight (w) & Chiral weight (c) \\ \hline
			$G$ & $\bf{1}$ & Complex & $3$ & $1$ \\ \hline
			$L_{ij}$ & $\bf{3}$ & $L^{ij}\equiv (L_{ij})^*=\varepsilon_{ik}\varepsilon_{jl}L^{kl}$ &2 & 0  \\ \hline
			$E_{\mu\nu}$ & $1$ & $E_{\mu\nu}=-E_{\nu\mu}$ Two-form gauge field &$0$ & $0$  \\ \hline
			$\phi^i$ & $\bf{2}$ & $\gamma_{5}\phi^i=\phi^i$ &$5/2$ & $1/2$  \\ \hline
		\end{tabular}
	\end{center}
\end{table} 

We will present the field content and the transformation rule of the $8+8$ tensor multiplet which is of relevance to this paper. Field content of the $\mathcal{N}=2$ tensor multiplet is given in Table-\ref{Table-Tensor}. 

All the fields are invariant under $K$-transformations. The component fields transform under $Q$ and $S$ supersymmetry is as follows.
\begin{align}\label{tensor_susy}
\delta L_{ij}&=2\bar{\epsilon}_{(i}\varphi_{j)}+2\varepsilon_{ik}\varepsilon_{jl}\bar{\epsilon}^{(k}\varphi^{l)} \;, \nonumber \\
\delta\varphi^{i}&=\slashed{D}L^{ij}\epsilon_{j}+\slashed{H}\varepsilon^{ij}\epsilon_{j}-G\epsilon^{i}+2L^{ij}\eta_{j}\;, \nonumber \\
\delta G&=-2\bar{\epsilon}_{i}\slashed{D}\varphi^{i}-6\bar{\epsilon}_{i}\chi_{j}L^{ij}+\frac{1}{4}\varepsilon^{ij}\bar{\epsilon}_{i}\gamma\cdot T^{+}\varphi_{j}+2\bar{\eta}_{i}\varphi^{i}\;, \nonumber \\
\delta E_{\mu\nu}&=i\bar{\epsilon}^{i}\gamma_{\mu\nu}\varphi^{j}\varepsilon_{ij}+2iL_{ij}\varepsilon^{jk}\bar{\epsilon}^{i}\gamma_{[\mu}\psi_{\nu]k}+\text{h.c}\;,
\end{align}
where $H^a$ is defined to be
\begin{align}
H^a=\frac{i}{6}\varepsilon_{abcd}H^{bcd}\;,
\end{align}
where $H_{\mu\nu\rho}$ is the fully supercovariant field strength corresponding to the two-form gauge field.

\section{Embedding of real scalar multiplet in the density formula: Intermediate results}\label{inter}
In this appendix, we will give all the intermediate results that we get in obtaining the final bosonic Lagrangian (\ref{R4}). By taking a left-supersymmetry transformation of $\Sigma_{ijk}$ (\ref{R1}) up to terms quadratic in fermions, we obtain:
\begin{align}\label{I}
\mathcal{G}^a_{ij}&={24iD^{(a}D^{b)}S_{bij}}-16iD^2S^a_{ij}+16iD_bR(V)^{ab}{}_{ij}-32i\varepsilon^{kl}D^a\bar{E}_{k(i}E_{j)l}\nonumber\\
&\;\;\;+32i\varepsilon^{kl}\bar{E}_{k(i}D^aE_{j)l}+16iD_bS^{bk}{}_{_(i}S^a_{j)k}-32iD_bS^{ak}{}_{(i}S^b_{j)k}+16iD^aS_{b}{}^k{}_{(i}S^b_{j)n}\nonumber\\
&\;\;\;-16iT^{+ab}D_bE_{ij}-16iT^{-ab}D_b\bar{E}_{ij}-16iD_bT^{-ab}\bar{E}_{ij}-16iD_bT^{+ab}E_{ij}+16iD^{(a}P^{b)}S_{bij}\nonumber\\
&\;\;\;+\frac{32i}{3}D_bP^bS^a_{ij}+16iP_bD^aS^b_{ij}-16iD_bS^b_{ij}P^a {+8\tilde{R}^{ab}(A)S_{bij}}+20iS_{bk(i}R(V)^{ab}{}^k{}_{j)}{+10iDS^a_{ij}}\nonumber\\
&\;\;\;-\frac{8i}{3}S^{akl}C_{ijkl}-32iP_bS^b_{ij}P^a+\frac{32i}{3}P^bP_bS^a_{ij}+\frac{32i}{3}S^{akl}E_{kl}\bar{E}_{ij}+\frac{32i}{3}S^{akl}\bar{E}_{kl}E_{ij}\nonumber \\
&\;\;\;-\frac{160i}{3}S^{akl}E_{k(i}\bar{E}_{j)l}+\frac{2i}{3}S^a_{ij}S^{bkl}S_{bkl}-\frac{4i}{3}S^{akl}S_{bkl}S^b_{ij}-\frac{4i}{3}S^{akl}S_{bk(i}S^b_{j)l}\nonumber\\
&\;\;\; +iT^{-}{}^{ac}T^{+}_{bc}S^{b}_{ij}-\frac{20i}{9}D^a\bar{\Lambda}^k\slashed{D}\Lambda_{(i}\varepsilon_{j)k}-\frac{64i}{9}D_b\bar{\Lambda}^k\gamma^{ab}\slashed{D}\Lambda_{(i}\varepsilon_{j)k}-\frac{28i}{3}D_b\bar{\Lambda}^k\gamma^bD^a\Lambda_{(i}\varepsilon_{j)k}\nonumber\\
&\;\;\;+16iD_b\bar{\Lambda}^k\gamma^aD^b\Lambda_{(i}\varepsilon_{j)k}-8iD^a\bar{\Lambda}^k\chi_{(i}\varepsilon_{j)k}-\frac{46i}{3}D_b\bar{\Lambda}^k\gamma^b\gamma^a\chi_{(i}\varepsilon_{j)k}+8iD^a\bar{\Lambda}_{(i}\varepsilon_{j)k}\chi^k\nonumber\\
&\;\;\;+\frac{46i}{3}\bar{\chi}^k\gamma^a\slashed{D}\Lambda_{(i}\varepsilon_{j)k}+\frac{16i}{9}\varepsilon^{kl}\bar{\Xi}_{ijk}\gamma^a\slashed{D}\Lambda_l +\frac{8i}{3}\varepsilon^{kl}D^a\bar{\Lambda}_l\Xi_{ijk}+\frac{8i}{3}D^a\bar{\Lambda}^n\Xi^{klm}\varepsilon_{ik}\varepsilon_{jl}\varepsilon_{mn}\nonumber\\
&\;\;\;+\frac{16i}{{9}}\bar{\Xi}^{klm}\gamma^a\slashed{D}\Lambda^n\varepsilon_{ik}\varepsilon_{jl}\varepsilon_{mn}+\frac{25i}{3}\varepsilon^{kl}\bar{\chi}_l\gamma^a\Xi_{ijk}-\frac{25i}{3}\bar{\chi}^k\gamma^a\Xi^{lmn}\varepsilon_{kl}\varepsilon_{im}\varepsilon_{jn}\nonumber\\
&\;\;\;+20i\bar{R}^{ab}(Q)_{(i}\varepsilon_{j)k}D_b\Lambda^k {-20iD_b\bar{\Lambda}_{(i}\varepsilon_{j)k}R^{ab}(Q)^k} -42i\bar{\chi}^k\gamma^a\chi_{(i}\varepsilon_{j)k}+\frac{16i}{9}\bar{\Xi}^{klm}\gamma^a\Xi_{kl(i}\varepsilon_{j)m}\nonumber \\
&\;\;\; -\frac{10i}{3}\varepsilon^{kl}\bar{\Lambda}_l\gamma^a\slashed{D}\Xi_{ijk}+\frac{8i}{3}\varepsilon^{kl}\bar{\Lambda}_lD^a\Xi_{ijk}-\frac{8i}{3}\bar{\Lambda}^kD^a\Xi^{lmn}\varepsilon_{kl}\varepsilon_{im}\varepsilon_{jn}-\frac{14i}{3}\bar{\Lambda}^k\gamma^a\slashed{D}\Xi^{lmn}\varepsilon_{kl}\varepsilon_{im}\varepsilon_{jn}\nonumber\\
&\;\;\;+8i\bar{\Lambda}_{(i}\varepsilon_{j)k}D^a\chi^k-13i\bar{\Lambda}_{(i}\varepsilon_{j)k}\gamma^a\slashed{D}\chi^k-8i\bar{\Lambda}^kD^a\chi_{(i}\varepsilon_{j)k}+13i\bar{\Lambda}^k\gamma^a\slashed{D}\chi_{(i}\varepsilon_{j)k} \nonumber\\
&\;\;\; +\frac{68i}{3}\bar{\Lambda}_{(i}\varepsilon_{j)k}D_bR^{ba}(Q)^k -\frac{68i}{3}\bar{\Lambda}^kD_{b}R^{ba}(Q)_{(i}\varepsilon_{j)k}-\frac{8i}{3}\bar{\Lambda}_{(i}\varepsilon_{j)k}\gamma^aD^2\Lambda^k +\frac{28i}{3}\bar{\Lambda}_{(i}\varepsilon_{j)k}\gamma_bD^{(a}D^{b)}\Lambda^k \nonumber\\
&\;\;\;+\frac{8i}{3}\bar{\Lambda}^k\gamma^aD^2\Lambda_{(i}\varepsilon_{j)k}-\frac{28i}{3}\bar{\Lambda}^k\gamma_bD^{(a}D^{b)}\Lambda_{(i}\varepsilon_{j)k}    +\frac{16i}{9}\bar{\Lambda}_l\slashed{P}\gamma^a\Xi_{ijk}\varepsilon^{kl}-\frac{16i}{3}\bar{\Lambda}_l\Xi_{ijk}P^a\varepsilon^{kl}\nonumber\\
&\;\;\;-\frac{16i}{9}\bar{\Lambda}^k\slashed{P}\gamma^a\Xi^{lmn}\varepsilon_{kl}\varepsilon_{im}\varepsilon_{jn}+\frac{16i}{3}\bar{\Lambda}^k\Xi^{lmn}P^a\varepsilon_{kl}\varepsilon_{im}\varepsilon_{jn}+\frac{8i}{3}\varepsilon^{lm}\bar{\Lambda}^k\gamma^a\Xi_{kl(i}\bar{E}_{j)m}\nonumber\\
&\;\;\;+\frac{44i}{9}\varepsilon^{kl}\bar{\Lambda}^m\gamma^a\Xi_{ijk}\bar{E}_{lm}+\frac{4i}{9}\bar{\Lambda}_{(i}\varepsilon_{j)k}\gamma^a\Xi^{klm}E_{lm}-\frac{76i}{9}\bar{\Lambda}_k\gamma^a\Xi^{klm}E_{l(i}\varepsilon_{j)m}-\frac{8i}{3}\varepsilon^{mn}\varepsilon^{kl}\bar{\Lambda}_k\Xi_{lm(i}S^a_{j)n}\nonumber\\
&\;\;\;-4i\bar{\Lambda}_l\Xi_{ijk}S^{akl}+\frac{4i}{3}\varepsilon^{kl}\bar{\Lambda}_l\slashed{S}{}^m{}_{(i}\gamma^a\Xi_{j)lm}+\frac{10i}{9}\bar{\Lambda}_l\slashed{S}^{kl}\gamma^a\Xi_{ijk}+8i\bar{\Lambda}^k\Xi^{lmn}\varepsilon_{kl}S^a_{m(i}\varepsilon_{j)n}\nonumber\\
&\;\;\;+\frac{4i}{3}\bar{\Lambda}^k\Xi^{lmn}S^a_{kl}\varepsilon_{im}\varepsilon_{jn}+4i\bar{\Lambda}^n\slashed{S}_{k(i}\varepsilon_{j)l}\gamma^a\Xi^{klm}\varepsilon_{mn}-\frac{14i}{9}\bar{\Lambda}^k\slashed{S}_{kl}\gamma^a\Xi^{lmn}\varepsilon_{im}\varepsilon_{jn}-{\frac{8i}{9}}\bar{\Lambda}^k\gamma_b\Xi_{ijk}T^{+ab}\nonumber\\
&\;\;\;-{\frac{8i}{9}}\bar{\Lambda}_k\gamma_b\Xi^{klm}T^{-ab}\varepsilon_{il}\varepsilon_{jm}+\frac{16i}{3}\bar{\Lambda}^k\slashed{D}\gamma^{ab}\Lambda_{(i}\varepsilon_{j)k}P_b-\frac{64i}{9}\bar{\Lambda}^k\gamma^{ab}\slashed{D}\Lambda_{(i}\varepsilon_{j)k}P_b-\frac{40i}{3}\bar{\Lambda}^k\slashed{P}D^a\Lambda_{(i}\varepsilon_{j)k}\nonumber\\
&\;\;\;+\frac{136i}{9}\bar{\Lambda}^k\slashed{D}\Lambda_{(i}\varepsilon_{j)k}P^a-\frac{16i}{3}\bar{\Lambda}_{(i}\varepsilon_{j)k}\slashed{D}\gamma^{ab}\Lambda^kP_b+\frac{64i}{9}\bar{\Lambda}_{(i}\varepsilon_{j)k}\gamma^{ab}\slashed{D}\Lambda^kP_b+\frac{40i}{3}\bar{\Lambda}_{(i}\varepsilon_{j)k}\slashed{P}D^a\Lambda^k\nonumber\\
&\;\;\;-\frac{136i}{9}\bar{\Lambda}_{(i}\varepsilon_{j)k}\slashed{D}\Lambda^kP^a+2i\bar{\Lambda}^k\gamma^a\slashed{D}\Lambda^l\bar{E}_{l(i}\varepsilon_{j)k}-\frac{38i}{9}\bar{\Lambda}^k\gamma^a\slashed{D}\Lambda^l\bar{E}_{k(i}\varepsilon_{j)l}+\frac{8i}{3}\bar{\Lambda}^kD^a\Lambda^l\bar{E}_{l(i}\varepsilon_{j)k}\nonumber\\
&\;\;\;+\frac{8i}{3}\bar{\Lambda}^kD^a\Lambda^l\bar{E}_{k(i}\varepsilon_{j)nl}+\frac{110i}{9}\bar{\Lambda}_l\gamma^a\slashed{D}\Lambda_{(i}E_{j)k}\varepsilon^{kl}-\frac{14i}{3}\bar{\Lambda}_{(i}E_{j)k}\gamma^a\slashed{D}\Lambda_l\varepsilon^{kl}\nonumber\\
&\;\;\; -\frac{8i}{3}\bar{\Lambda}_lD^a\Lambda_{(i}E_{j)k}\varepsilon^{kl}-\frac{8i}{3}\bar{\Lambda}_{(i}E_{j)k}D^a\Lambda_l\varepsilon^{kl}+\frac{16i}{3}\bar{\Lambda}^k\slashed{D}\gamma^a{}_b\Lambda_{(i}S^b_{j)k}+\frac{20i}{9}\bar{\Lambda}^k\gamma^a{}_b\slashed{D}\Lambda_{(i}S^b_{j)k}\nonumber \\
&\;\;\;+\frac{76i}{9}\bar{\Lambda}^k\slashed{D}\Lambda_{(i}S^a_{j)k}-\frac{16i}{3}\bar{\Lambda}^k\slashed{S}_{k(i}D^a\Lambda_{j)}-\frac{4i}{3}\bar{\Lambda}^k\slashed{D}\gamma^a{}_b\Lambda_kS^b_{ij}+2i\bar{\Lambda}^k\gamma^a{}_b\slashed{D}\Lambda_kS^b_{ij}-\frac{2i}{3}\bar{\Lambda}^k\slashed{D}\Lambda_kS^a_{ij}\nonumber\\
&\;\;\;+\frac{28i}{9}\bar{\Lambda}_{(i}S^b_{j)k}\gamma^a{}_b\slashed{D}\Lambda^k-\frac{16i}{3}\bar{\Lambda}_{(i}S^b_{j)k}\slashed{D}\gamma^a{}_b\Lambda^k-\frac{28i}{9}\bar{\Lambda}_{(i}S^a_{j)k}\slashed{D}\Lambda^k+\frac{16i}{3}\bar{\Lambda}_{(i}\slashed{S}_{j)k}D^a\Lambda^k\nonumber \\
&\;\;\; -\frac{34i}{9}\bar{\Lambda}_k\gamma^{ab}\slashed{D}\Lambda^kS_{bij}+4i\bar{\Lambda}_k\slashed{D}\gamma^{ab}\Lambda^kS_{bij}-\frac{2i}{9}\bar{\Lambda}_k\slashed{D}\Lambda^kS^a_{ij}-\frac{16i}{3}\bar{\Lambda}_k\slashed{S}_{ij}D^a\Lambda^k-\frac{8i}{9}\bar{\Lambda}_{(i}\gamma^b\slashed{D}\Lambda_{j)}T^{-a}{}_b\nonumber\\
&\;\;\;-\frac{16i}{3}\bar{\Lambda}_{(i}D^b\Lambda_{j)}T^{-a}{}_b +\frac{4i}{3}\bar{\Lambda}_{(i}\gamma\cdot T^-D^a\Lambda_{j)}+\frac{8i}{9}\bar{\Lambda}^k\gamma_b\slashed{D}\Lambda^l T^{+ab}\varepsilon_{k(i}\varepsilon_{j)l}\nonumber\\
&\;\;\; -\frac{4i}{3}\bar{\Lambda}^k\gamma\cdot T^+D^a\Lambda^l\varepsilon_{k(i}\varepsilon_{j)l}+\frac{16i}{3}\bar{\Lambda}^kD_b\Lambda^lT^{+ab}\varepsilon_{k(i}\varepsilon_{j)l}+\frac{4i}{3}\bar{\Lambda}_{(i}\chi^{k}\varepsilon_{j)k}P^a-\frac{52i}{3}\bar{\Lambda}_{(i}\gamma^{ab}\chi^{k}\varepsilon_{j)k}P_b\nonumber \\
&\;\;\; -\frac{4i}{3}\bar{\Lambda}^{k}\chi_{(i}\varepsilon_{j)k}P^a+\frac{52i}{3}\bar{\Lambda}^{k}\gamma^{ab}\chi_{(i}\varepsilon_{j)k}P_b -\frac{8i}{3}\bar{\Lambda}_{(i}R(Q)^{ab}{}^{k}\varepsilon_{j)k}P_b+\frac{8i}{3}\bar{\Lambda}^{k}R(Q)^{ab}{}_{(i}\varepsilon_{j)k}P_b \nonumber \\
&\;\;\;+11i\bar{\Lambda}_{k}\gamma^{a}\chi_{l}E_{ij}\varepsilon^{kl} -\frac{22i}{3}\bar{\Lambda}_{k}\gamma^{a}\chi_{(i}E_{j)l}\varepsilon^{kl}+27i\bar{\Lambda}^{k}\gamma^{a}\chi^{l}\bar{E}_{ij}\varepsilon_{kl}+\frac{74i}{3}\bar{\Lambda}^{k}\gamma^{a}\chi^{l}\bar{E}_{k(i}\varepsilon_{j)l} \nonumber \\
&\;\;\;-\frac{9i}{4}\bar{\Lambda}_{(i}\gamma\cdot T^{-}\gamma^{a}\chi_{j)}-\frac{9i}{4}\bar{\Lambda}^{(k}\gamma\cdot T^{+}\gamma^{a}\chi^{l)}\varepsilon_{ik}\varepsilon_{jl}-\frac{14i}{3}\bar{\Lambda}_{(i}\chi^{k}S^{a}_{j)k}+\frac{38i}{3}\bar{\Lambda}_{(i}\gamma^{ab}\chi^{k}S_{b}{}_{j)k}{+\frac{29i}{3}}\bar{\Lambda}_{k}\chi^{k}S^{a}_{ij} \nonumber \\
&\;\;\; +\frac{7i}{3}\bar{\Lambda}_{k}\gamma^{ab}\chi^{k}S_{b}{}_{ij}-\frac{34i}{3}\bar{\Lambda}^{k}\chi_{(i}S^{a}_{j)k}-\frac{86i}{3}\bar{\Lambda}^{k}\gamma^{ab}\chi_{(i}S_{b}{}_{j)k}-3i\bar{\Lambda}^{k}\chi_{k}S^{a}_{ij}+7i\bar{\Lambda}^{k}\gamma^{ab}\chi_{k}S_{b}{}_{ij} \nonumber \\
&\;\;\; +\frac{64i}{3}\bar{\Lambda}_{(i}R(Q)^{ab}{}^{k}S_{b}{}_{j)k}-8i\bar{\Lambda}_{k}R(Q)^{ab}{}^{k}S_{b}{}_{ij}-\frac{64i}{3}\bar{\Lambda}^{k}R(Q)^{ab}_{(i}S_{b}{}_{j)k}+\frac{40i}{3}\bar{\Lambda}^{k}R(Q)^{ab}_{k}S_{b}{}_{ij} \nonumber \\
&\;\;\;+\frac{8i}{3}\bar{\Lambda}_{i}\Lambda_{j}D_b T^{-}{}^{ba}+\frac{8i}{3}\bar{\Lambda}^{k}\Lambda^{l}D_b T^{+}{}^{ba}\varepsilon_{ik}\varepsilon_{jl}-\frac{16i}{3}\bar{\Lambda}^l\gamma^a\Lambda_{(i}\varepsilon_{j)l}D^bP_b-\frac{8i}{3}\bar{\Lambda}^l\gamma_b\Lambda_{(i}\varepsilon_{j)l}D^{(a}P^{b)}
\nonumber \\
&\;\;\;+\frac{20i}{3}\varepsilon_{mn}\bar{\Lambda}^m\gamma^{ab}\Lambda^nD_b\bar{E}_{ij}+\frac{52i}{3}\bar{\Lambda}^m\Lambda^nD^a\bar{E}_{m(i}\varepsilon_{j)n}-8i\varepsilon^{mn}\bar{\Lambda}_m\gamma^{ab}\Lambda_nD_bE_{ij}\nonumber\\
&\;\;\;+\frac{8i}{3}\varepsilon^{mn}\bar{\Lambda}_m\Lambda_{(i}D^aE_{j)n}-\frac{16i}{3}\bar{\Lambda}^l\gamma_b\Lambda_{(i}D^aS^b_{j)l}+\frac{32i}{3}\bar{\Lambda}\gamma_b\Lambda_{(i}D^bS^a_{j)l}-\frac{20i}{3}\bar{\Lambda}^l\gamma^aD^bS_{bl(i}\Lambda_{j)}\nonumber\\
&\;\;\;+\frac{8i}{3}\bar{\Lambda}^l\gamma^{ab}\gamma^c\Lambda_lD_cS_{bij}+\frac{16i}{3}\bar{\Lambda}^l\gamma_b\Lambda_lD^aS^b_{ij}-\frac{16i}{3}\bar{\Lambda}^l\gamma_b\Lambda_lD^bS^a_{ij}-\frac{22i}{3}\bar{\Lambda}^l\gamma^a\Lambda_lD^bS_{bij}\nonumber\\
&\;\;\;+\frac{8i}{3}\varepsilon^{lm}\bar{\Lambda}^n\gamma^a\Lambda_lC_{ijmn} -\frac{17i}{12}\bar{\Lambda}^{k}\gamma\cdot R(V)_{ij}\gamma^{a}\Lambda_{k}+\frac{i}{12}\bar{\Lambda}^{k}\gamma^{a}\gamma\cdot R(V)_{ij}\Lambda_{k}+\frac{3i}{2}\bar{\Lambda}^{k}\gamma\cdot R(V)_{k(i}\gamma^{a}\Lambda_{j)} \nonumber \\
&\;\;\; -\frac{3i}{2}\bar{\Lambda}^{k}\gamma^{a}\gamma\cdot R(V)_{k(i}\Lambda_{j)}+\frac{1}{3}\bar{\Lambda}^{k}\gamma\cdot R(A)\gamma^{a}\Lambda_{(i}\varepsilon_{j)k}+\frac{1}{3}\bar{\Lambda}^{k}\gamma^{a}\gamma\cdot R(A)\Lambda_{(i}\varepsilon_{j)k}-3i\bar{\Lambda}^{k}\gamma^{a}\Lambda_{(i}\varepsilon_{j)k}D \nonumber \\
&\;\;\; +\frac{11i}{6}\bar{\Lambda}^{k}\gamma^{b}\Lambda_{(i}\varepsilon_{j)k}T^{-}{}^{af}T^{+}_{bf}-\frac{32i}{9}\bar{\Lambda}_{i}\Lambda_{j}P_{b}T^{-}{}^{ab}-\frac{32i}{9}\bar{\Lambda}^{k}\Lambda^{l}P_{b}T^{+}{}^{ab}\varepsilon_{ik}\varepsilon_{jl}+\frac{26i}{9}\bar{\Lambda}^{k}\gamma_{b}\Lambda_{k}E_{ij}T^{+}{}^{ab} \nonumber \\
&\;\;\; +\frac{44i}{9}\bar{\Lambda}^{k}\gamma_{b}\Lambda_{(i}E_{j)k}T^{+}{}^{ab}-\frac{14i}{3}\bar{\Lambda}^{k}\gamma_{b}\Lambda_{k}\bar{E}_{ij}T^{-}{}^{ab}-\frac{4i}{9}\bar{\Lambda}^{k}\gamma_{b}\Lambda_{(i}\bar{E}_{j)k}T^{-}{}^{ab} +\frac{22i}{9}\bar{\Lambda}_{k}\Lambda_{(i}S_{b}{}_{j)l}T^{-}{}^{ab}\varepsilon^{kl}\nonumber \\
&\;\;\; +\frac{46i}{9}\bar{\Lambda}^{k}\Lambda^{l}S_{b}{}_{k(i}\varepsilon_{j)l}T^{+}{}^{ab}+\frac{10i}{9}\bar{\Lambda}_{k}\gamma^{ac}\Lambda_{l}S^{b}_{ij}T^{-}_{bc}\varepsilon^{kl}+\frac{i}{9}\bar{\Lambda}_{k}\gamma\cdot T^{-}\Lambda_{l}S^{a}_{ij}\varepsilon^{kl} \nonumber \\
&\;\;\; -\frac{14i}{9}\bar{\Lambda}^{k}\gamma^{ac}\Lambda^{l}S^{b}_{ij}T^{+}_{bc}\varepsilon_{kl}+\frac{i}{9}\bar{\Lambda}^{k}\gamma\cdot T^{+}\Lambda^{l}S^{a}{}_{ij}\varepsilon_{kl}-\frac{16i}{9}\bar{\Lambda}^l\gamma^a\Lambda_{(i}\varepsilon_{j)l}P^bP_b+\frac{176i}{9}\bar{\Lambda}^l\slashed{P}\Lambda_{(i}\varepsilon_{j)l}P^a\nonumber \\
&\;\;\;-\frac{88i}{9}\varepsilon_{kl}\bar{\Lambda}^k\gamma^{ab}\Lambda^l\bar{E}_{ij}P_b -\frac{200i}{9}\bar{\Lambda}^k\Lambda^l\bar{E}_{k(i}\varepsilon_{j)l}+\frac{8i}{9}\varepsilon^{kl}\bar{\Lambda}_k\gamma^{ab}\Lambda_lP_bE_{ij}\nonumber \\
&\;\;\; -\frac{104i}{9}\bar{\Lambda}_k\Lambda_{(i}E_{j)l}\varepsilon^{kl}P^a +\frac{16i}{3}\bar{\Lambda}^k\gamma^a{}_b\slashed{P}\Lambda_{(i}S^b_{j)k}-\frac{40i}{9}\bar{\Lambda}^k\slashed{S}_{k(i}\Lambda_{j)}P^a+\frac{8i}{9}\bar{\Lambda}^k\slashed{P}\Lambda_{(i}S^a_{j)k}\nonumber\\
&\;\;\;+\frac{16i}{9}\bar{\Lambda}^k\gamma^a\Lambda_{(i}S^b_{j)k}P_b -\frac{16i}{9}\bar{\Lambda}^k\gamma^a{}_b\slashed{P}\Lambda_kS_{bij}-\frac{20i}{9}\bar{\Lambda}^k\slashed{S}_{ij}\Lambda_kP^a+\frac{44i}{9}\bar{\Lambda}^k\slashed{P}\Lambda_kS^a_{ij}\nonumber \\
&\;\;\; -\frac{112i}{9}\bar{\Lambda}^k\gamma^a\Lambda_kP^bS_{bij}-\frac{10i}{9}\varepsilon_{kl}\Lambda^m\gamma^{ab}\Lambda^lS_b{}^m{}_{(i}\bar{E}_{j)m}-\frac{4i}{3}\bar{\Lambda}^k\Lambda^lS^a_{kl}\bar{E}_{ij}+\frac{10i}{3}\bar{\Lambda}^k\Lambda^l\bar{E}_{kl}S^a_{ij}\nonumber\\
&\;\;\;-\frac{44i}{9}\bar{\Lambda}^k\Lambda^lS^a_{k(i}\bar{E}_{j)l}-\frac{62i}{9}\varepsilon^{kl}\bar{\Lambda}_k\gamma^{ab}\Lambda_lS_b{}^m{}_{(i}E_{j)m}\varepsilon^{kl}-\frac{16i}{9}\bar{\Lambda}_{(i}E_{j)k}\Lambda_lS^{akl}\nonumber\\
&\;\;\;+\frac{92i}{9}\bar{\Lambda}_k\Lambda_lS^{akl}E_{ij}+\frac{4i}{3}\bar{\Lambda}_i\Lambda_jS^{amn}E_{mn}-\frac{56i}{3}\varepsilon^{kl}\Lambda^m\gamma^a\Lambda_{(i}E_{j)k}\bar{E}_{lm}-\frac{32i}{9}\varepsilon^{kl}\bar{\Lambda}^m\gamma^a\Lambda_{l}\bar{E}_{m(i}E_{j)k}\nonumber\\
&\;\;\;+\frac{8i}{3}\varepsilon^{kl}\bar{\Lambda}^m\gamma^a\Lambda_{(i}\bar{E}_{j)k}E_{lm}-\frac{64i}{3}\varepsilon^{kl}\bar{\Lambda}^m\gamma^a\Lambda_{m}\bar{E}_{k(i}E_{j)l}-\frac{26i}{9}\bar{\Lambda}^k\gamma^a\gamma^{bc}\Lambda_lS_{bk(i}S_c{}^l{}_{j)}\nonumber \\
&\;\;\;-\frac{4i}{3}\bar{\Lambda}^k\gamma^a\gamma^{bc}\Lambda_lS_{bij}S_c{}^l{}_k +\frac{32i}{9}\bar{\Lambda}^k\gamma^a\Lambda_lS_{bk(i}S^{bl}{}_{j)}-2i\bar{\Lambda}^k\gamma^a\Lambda_lS_{bij}S^{bl}{}_k\nonumber\\
&\;\;\;+\frac{2i}{3}\bar{\Lambda}^k\slashed{S}_{kl}\Lambda_{(i}S^{al}{}_{j)}-\frac{8i}{9}\bar{\Lambda}^k\slashed{S}_{l(i}S^{al}{}_{j)}\Lambda_k -\frac{46i}{9}\bar{\Lambda}^k\slashed{S}_{k(i}S^{al}{}_{j)}\Lambda_l+\frac{10i}{3}\bar{\Lambda}^k\slashed{S}_{ij}\Lambda_lS^{al}{}_k
\end{align}
Upon taking a further supersymmetry transformation of $\mathcal{G}^{a}_{ij}$ and using (\ref{d4.2}), we get $\Psi_{i}$ up to terms linear in fermions as:
\begin{align}\label{I1}
\Psi_{i}&= 4iD^2\slashed{D}\Lambda^{k}\varepsilon_{ik}-12iD^2\chi^{k}\varepsilon_{ik}+\frac{8i}{3}D^2 \Lambda_{k}E_{il}\varepsilon^{kl}+i \gamma\cdot T^{-}D^2 \Lambda_{i}-\frac{4i}{3}\slashed{S}_{ij}D^2\Lambda^{j}\nonumber \\
&\;\;\; -\frac{16i}{3}\gamma^a{S}^{b}_{ij}D_{(a}D_{b)}\Lambda^{j}+\frac{4i}{3}\slashed{D}\Xi^{jkl}E_{kl}\varepsilon_{ij}-\frac{4i}{3}D_{a}\Xi_{ijk}S^{a}{}^{jk}+\frac{5i}{12}\gamma_{a}\slashed{D}\Xi_{ijk}S^{a}{}^{jk}\nonumber \\
&\;\;\; -\frac{8i}{9}\slashed{D}E_{kl}\Xi^{jkl}\varepsilon_{ij} -\frac{5i}{9}\Xi_{ijk}D_{a}S^{a}{}^{jk}-\frac{2i}{3}\gamma^{ab}\Xi_{ijk}D_a S_{b}{}^{jk}-\frac{4i}{9}\Xi_{jkl}C^{jklm}\varepsilon_{im}\nonumber \\
& \;\;\; -i\gamma\cdot R(V)_{ij}\slashed{D}\Lambda^{j}-\gamma\cdot R(A)\slashed{D}\Lambda^{k}\varepsilon_{ik}+\gamma^a \gamma\cdot R(A)D_a \Lambda^{k}\varepsilon_{ik}-6i\slashed{D}\Lambda^{k}\varepsilon_{ik}D\nonumber \\
&\;\;\; -8i\slashed{D}\Lambda^{k}D^{b}P_{b}\varepsilon_{ik}+8i\gamma^{a}D^{b}\Lambda^{k}D_{(a}P_{b)}\varepsilon_{ik}+\frac{24 i}{9}\varepsilon^{kl}D^{a}\Lambda_{k}D_{a}E_{il}+\frac{48 i}{9}\varepsilon^{kl}\gamma^{ab}D_{a}\Lambda_{k}D_{b}E_{il}\nonumber \\
& \;\;\; -\frac{8i}{3}\slashed{D}\Lambda^{k}D^b S_{bik}-\frac{8i}{3}\gamma^{ab}\slashed{D}\Lambda^{k}D_{a} S_{bik}-4iD^{a}\slashed{S}_{ik}D_{a}\Lambda^{k}+i\gamma\cdot D_{b}T^{-}D^{b}\Lambda_{i}\nonumber \\
& \;\;\; +6i\slashed{P}\slashed{D}\chi^j\varepsilon_{ij}+4i\varepsilon^{jk}E_{ij}\slashed{D}\chi_k-5i\slashed{S}_{ij}\slashed{D}\chi^j+16iS^b_{ij}D_b\chi^j+\frac{16i}{3}S^b_{ij}D^aR_{ab}(Q)^j\nonumber\\
&\;\;\;-\frac{3i}{4}\gamma\cdot T^-\slashed{D}\chi_i+12iD_bP^b\varepsilon_{ij}\chi^j+14i\gamma^{ab}D_aS_{bij}\chi^j+14iD_bS^b_{ij}\chi^j+8iD_aS_{bij}R^{ab}(Q)^j\nonumber\\
&\;\;\; -16i\varepsilon^{jk}\slashed{D}E_{ij}\chi_k-\frac{i}{2}\gamma\cdot\slashed{D}T^-\chi_i +4iR(V)_{ab}{}_{ij}R^{ab}(Q)^j +6\varepsilon_{ij}\gamma\cdot R(A)\chi^j\nonumber\\
&\;\;\; +12i D\varepsilon_{ij}\chi^j-iD^2 S_{aij}\gamma^{a}\Lambda^{j}-\frac{i}{2}D_{(a}D_{b)}S^{b}_{ij}\gamma^{a}\Lambda^{j}+\frac{i}{2}\gamma\cdot D^2 T^{-}\Lambda_{i}+2\gamma\cdot\slashed{D}R(A)\Lambda^{j}\varepsilon_{ij}\nonumber \\
 & \;\;\; +\frac{i}{4}\gamma\cdot\slashed{D}R(V)_{ij}\Lambda^{j}+\frac{3i}{4}\slashed{D}\gamma\cdot R(V)_{ij}\Lambda^{j}-6i\slashed{D}(D)\Lambda^{k}\varepsilon_{ik}-4i\slashed{D}\Lambda^{k}\varepsilon_{ik}P^a P_a\nonumber \\
&\;\;\;-8i \gamma^a D^b \Lambda^{k}\varepsilon_{ik}P_a P_b-\frac{8i}{3}\gamma^{ab}D_{a}\Lambda_{k}E_{jl}P_{b}\varepsilon^{kl}-\frac{4i}{3}\varepsilon^{abcd}\gamma_{a}D_{b}\Lambda^{k}P_{c}S_{dik} +4i P^{a}D^{b}\Lambda_{i}T^{-}_{ab}\nonumber \\
&\;\;\; +\frac{5i}{9}\slashed{D}\Lambda^{p}\bar{E}_{il}E_{kp}\varepsilon^{kl}+\frac{31i}{9}\slashed{D}\Lambda^{p}{E}_{il}\bar{E}_{kp}\varepsilon^{kl}-\frac{3i}{2}D_{a}\Lambda_{l}S^{a}_{ik}\bar{E}^{kl}+\frac{11i}{2}\gamma^{ab}D_{a}\Lambda_{l}S_{b}{}_{ik}\bar{E}^{kl}\nonumber\\
& \;\;\;+\frac{17i}{4}D_{a}\Lambda_{i}S^{a}_{kl}\bar{E}^{kl}-\frac{19i}{12}\gamma^{ab}D_{a}\Lambda_{i}S_{b}{}_{kl}\bar{E}^{kl}+\frac{i}{6}\slashed{D}\Lambda^{k}S_{a}{}^{l}{}_{k}S^{a}_{il}+\frac{2i}{3}\gamma^{ab}\slashed{D}\Lambda^{k}S_{a}{}^{l}{}_{k}S_{b}{}_{il}\nonumber\\
&\;\;\;-\frac{5i}{6}\slashed{S}^{l}{}_{k}D_{a}\Lambda^{k}S^{a}_{il}-\frac{i}{6}\gamma^{ab}\slashed{S}^{l}{}_{k}D_{a}\Lambda^{k}S_{b}{}_{il}-i\gamma^a D^{b}\Lambda^{j}T^{-}{}^{c}{}_{a}T^{+}_{cb}\varepsilon_{ij}-\frac{i}{3}\gamma^{a}\gamma\cdot T^{+} D_{a}\Lambda^{j}E_{ij}\nonumber\\
&\;\;\;-\frac{5i}{6}\gamma^{a}\slashed{S}^{j}{}_{i}\gamma\cdot T^{-}D_{a}\Lambda_{j}-\frac{i}{12}\gamma\cdot T^{-}\slashed{S}^{j}{}_{i}\slashed{D}\Lambda_{j}+\frac{i}{2}\gamma\cdot T^{-}D^a \Lambda_{j} {S}_{a}{}^{j}{}_{i}+12i\varepsilon_{ij}P^aP_a\chi^j \nonumber\\
&\;\;\;-\frac{11i}{2}\gamma^{ab}S_a{}^l{}_iS_{blm}\chi^m-6iS_a{}^l{}_iS_{alm}\chi^m-\frac{4i}{3}S^a{}^l{}_iS^b_{lm}R_{ab}(Q)^m-8i\varepsilon^{kl}E_{il}\bar{E}_{km}\chi^m \nonumber\\
&\;\;\;+4i\varepsilon^{kl}\slashed{P}E_{ik}\chi_l-4i\gamma^{ab}P_aS_{bij}\chi^j+8iP^aS_{aij}\chi^j+\frac{16i}{3}P_aS_{bij}R^{ab}(Q)^j-\frac{i}{2}\gamma\cdot T^-\slashed{P}\chi_i \nonumber\\
&\;\;\; +\frac{19i}{2}\slashed{S}^{mn}E_{mn}\chi_i -19i\slashed{S}^{mn}E_{im}\chi_n +\frac{13i}{12}\gamma\cdot T^-\slashed{S}^j{}_i\chi_j-3i\gamma\cdot T^-\bar{E}_{ij}\chi^j-2iT^{-ab}R_{ab}(Q)^j\bar{E}_{ij}\nonumber \\
&\;\;\; +\frac{8i}{9}\slashed{P}\varepsilon_{ij}E_{kl}\Xi^{jkl}+\frac{8i}{9}P^aS_a^{jk}\Xi_{ijk}+\frac{2i}{3}\gamma_{ab}P^aS^{bjk}\Xi_{ijk}+\frac{i}{24}\varepsilon_{il}\gamma\cdot T^-\slashed{S}_{mn}\Xi^{lmn}\nonumber\\
&\;\;\;+\frac{8i}{9}\varepsilon_{il}E^{lm}\bar{E}^{np}\Xi_{mnp}-\frac{44i}{9}\varepsilon_{il}\bar{E}^{lm}E^{np}\Xi_{mnp}+\frac{3i}{2}E_{il}\slashed{S}_{mn}\Xi^{lmn}-\frac{7i}{3}\slashed{S}_{il}E_{mn}\Xi^{lmn}\nonumber\\
&\;\;\;+\frac{i}{12}\gamma^{ab}S_a{}^l{}_iS_b^{mn}\Xi_{lmn}+\frac{i}{12}\gamma^{ab}S_a{}^m{}_lS_b^{ln}\Xi_{imn}+\frac{i}{12}S_a{}^{l}{}_iS_a^{mn}\Xi_{lmn}-4i \slashed{P}\Lambda^{l}\varepsilon_{il}D^{b}P_{b}\nonumber\\\
& \;\;\;-8i\gamma^{a}\Lambda^{l}\varepsilon_{il}P^{b}D_{(a}P_{b)}-\frac{28 i}{3}\varepsilon^{kl}\Lambda_{k}D^{a}E_{il} P_a-\frac{4 i}{3}\varepsilon^{kl}\gamma^{ab}\Lambda_{k}D_{a}E_{il} P_b-3i \slashed{D}S^{b}_{ij}\Lambda^{j}P_{b}\nonumber\\\
& \;\;\;-\frac{5i}{3} \slashed{P}\Lambda^{j}D_{b}S^{b}_{ij}-\frac{2i}{3} \gamma^{ab}\slashed{P}\Lambda^{j}D_{a}S_{b}{}_{ij}+4i {D}^{b}\slashed{S}_{ij}\Lambda^{j}P_{b}-\frac{i}{2}\slashed{P}\slashed{D}\gamma\cdot T^{-}\Lambda_{i}-\frac{20 i}{3}\varepsilon^{kl}\Lambda_{k}E_{il} D^{a}P_a\nonumber\\\
& \;\;\;+4i\slashed{S}_{ij}\Lambda^{j}D^{b}P_{b}-\frac{13i}{3}\gamma^a {S}^{b}_{ij}\Lambda^{j}D_{(a}P_{b)}-\frac{3i}{2}\gamma\cdot T^{-}\Lambda_{i}D^{a}P_{a}+\frac{5i}{3}\slashed{D}\bar{E}_{lk}\Lambda^{k}E_{im}\varepsilon^{lm}\nonumber\\\
& \;\;\;+i\slashed{D}\bar{E}_{im}\Lambda^{k}E_{lk}\varepsilon^{lm}+3i\slashed{D}{E}_{lk}\Lambda^{k}\bar{E}_{im}\varepsilon^{lm}+\frac{11i}{3}\slashed{D}{E}_{im}\Lambda^{k}\bar{E}_{lk}\varepsilon^{lm}+\frac{49i}{12}\Lambda_{i}S^{a}_{kl}D_{a}\bar{E}^{kl}\nonumber\\\
&\;\;\; -\frac{17i}{12}\gamma_{ab}\Lambda_{i}S^{b}_{kl}D^{a}\bar{E}^{kl}-\frac{5i}{2}\Lambda_{k}S^{a}_{il}D_{a}\bar{E}^{kl}+\frac{5i}{2}\gamma_{ab}\Lambda_{k}S^{b}_{il}D^{a}\bar{E}^{kl}+\frac{16i}{3}\Lambda_{i}D^{a}S_{a}{}_{jk}\bar{E}^{jk}\nonumber\\\
& \;\;\;+\frac{4i}{3}\gamma^{ab}\Lambda_{i}D_{a}S_{b}{}_{jk}\bar{E}^{jk}-2i\Lambda_{k}D^{a}S_{a}{}_{ij}\bar{E}^{jk}+2i\gamma^{ab}\Lambda_{k}D_{a}S_{b}{}_{ij}\bar{E}^{jk}+\frac{9i}{16}\slashed{D}S_{a}{}^{j}{}_{k}\Lambda^{k}S^{a}_{ij}\nonumber\\\
&\;\;\; +i\slashed{D}S^{a}_{ij}\Lambda^{k}S_{a}{}^{j}{}_{k}-\frac{i}{16}\gamma^{ab}\slashed{D}S_{a}{}^{j}{}_{k}\Lambda^{k}S_{b}{}_{ij}+\frac{i}{2}\gamma^{ab}\slashed{D}S_{b}{}_{ij}\Lambda^{k}S_{a}{}^{j}{}_{k}-\frac{61i}{48}D_{a}\slashed{S}^{j}{}_{k}\Lambda^{k}S^{a}_{ij}\nonumber\\
&\;\;\; -\frac{35i}{24}\slashed{S}^{j}{}_{k}\Lambda^{k}D_{a}S^{a}_{ij}-\frac{19i}{48}\gamma^{ab}D_{a}\slashed{S}^{j}{}_{k}\Lambda^{k}S_{b}{}_{ij}-\frac{5i}{6}\gamma^{ab}\slashed{S}^{j}{}_{k}\Lambda^{k}D_{a}S_{b}{}_{ij}-\frac{43i}{48}\gamma^{a}\Lambda^{j}D^{b}T^{-}{}_{a}{}^{c}T^{+}_{bc}\varepsilon_{ij}\nonumber\\
&\;\;\;-i\gamma^{a}\Lambda^{j}T^{-}{}_{a}{}^{c}D^{b}T^{+}_{bc}\varepsilon_{ij}-\frac{i}{4}\gamma\cdot T^{-}\slashed{D}\bar{E}_{ij}\Lambda^{j}+\frac{5i}{12}\slashed{D}{E}_{ij}\gamma\cdot T^{+}\Lambda^{j}+\frac{i}{4}\gamma\cdot\slashed{D}T^{-}\Lambda^{j}\bar{E}_{ij}\nonumber\\
& \;\;\;-\frac{i}{12}\slashed{D}\gamma\cdot T^{+}\Lambda^{j}{E}_{ij}+\frac{5i}{12}\gamma^{ab}\gamma\cdot T^{-}\Lambda_{j}D_{a}S_{b}{}^{j}{}_{i}-\frac{65i}{96}\gamma\cdot T^{-}\Lambda_{j}D^{b}S_{b}{}^{j}{}_{i}+\frac{25 i}{192}\slashed{S}^{j}{}_{i}\slashed{D}\gamma\cdot T^{-}\Lambda_{j}\nonumber \\
&\;\;\; -\frac{7 i}{64}\gamma\cdot \slashed{D}T^{-}\slashed{S}^{j}{}_{i}\Lambda_{j}-\frac{61i}{96}\gamma\cdot D^{a}T^{-}\Lambda_{j}{S}_{a}{}^{j}{}_{i} +i\slashed{P}\gamma\cdot R(V)_{ij}\Lambda^j-i\gamma\cdot R(V)_{ij}\slashed{P}\Lambda^j\nonumber \\
&\;\;\; -\varepsilon_{ij}\gamma\cdot R(A)\slashed{P}\Lambda^j+\varepsilon_{ij}\slashed{P}\gamma\cdot R(A)\Lambda^j -\frac{4i}{3}\gamma\cdot R(V)^{mn}E_{im}\Lambda_n+i\gamma\cdot R(V)^{mn}E_{mn}\Lambda_i \nonumber\\
&\;\;\;+2\gamma\cdot R(A)\varepsilon^{kl}E_{ik}\Lambda_l+4iD\varepsilon^{kl}E_{ik}\Lambda_l+\frac{13i}{48}\slashed{S}_{im}\gamma\cdot R(V)^{mn}\varepsilon_{nl}\Lambda^l-\frac{7i}{32}\slashed{S}_{mn}\gamma\cdot R(V)^{mn}\varepsilon_{il}\Lambda^l\nonumber\\&\;\;\;+\frac{19i}{48}\gamma\cdot R(V)^{mn}\slashed{S}_{im}\varepsilon_{nl}\Lambda^l-\frac{9i}{32}\gamma\cdot R(V)^{mn}\slashed{S}_{mn}\varepsilon_{il}\Lambda^l+\frac{13i}{12}\slashed{S}_{ij}\gamma\cdot R(A)\Lambda^j\nonumber\\
&\;\;\; -\frac{11i}{24}\gamma\cdot R(A)\slashed{S}_{ij}\Lambda^j+\frac{17i}{8}D\slashed{S}_{ij}\Lambda^j-\frac{i}{4}\varepsilon^{kl}\gamma\cdot R(V)_{ik}\gamma\cdot T^-\Lambda_l +\frac{i}{8}\varepsilon^{kl}\gamma\cdot T^-\gamma\cdot R(V)_{ik}\Lambda_l\nonumber\\
&\;\;\;+\gamma\cdot T^-\gamma\cdot R(A)\Lambda_i-\frac{i}{2}D\gamma\cdot T^-\Lambda_i+\frac{5i}{12}\slashed{S}^{jk}C_{ijkl}\Lambda^l+\frac{2i}{3}\varepsilon_{ij}C^{jklm}E_{kl}\Lambda_m \nonumber \\
&\;\;\; + \frac{2i}{3}\slashed{S}_{ij}P^2\Lambda^j+\frac{10i}{3}\slashed{P}P_aS^a_{ij}\Lambda^j+4i\varepsilon^{jk}E_{ij}\Lambda_kP^2-2iP^2\gamma\cdot T^-\Lambda_i +\frac{5i}{36}\slashed{P}\gamma^{ab}S_a{}^l{}_mS_{bim}\Lambda^m\nonumber\\
&\;\;\;+\frac{89i}{72}\slashed{P}S_a{}^l{}_mS^a_{il}\Lambda^m+\frac{35i}{72}\slashed{S}^l{}_mS^a_{il}P_a\Lambda^m -\frac{11i}{72}\slashed{S}_{il}S_a{}^l{}_mP^a\Lambda^m+\frac{3i}{2}\gamma_{ab}S^a_{mn}P^b\bar{E}^{mn}\Lambda_i\nonumber\\
&\;\;\;+\frac{65i}{6}P_aS^a_{mn}\bar{E}^{mn}\Lambda_i -i\gamma_{ab}S^a_{im}P^b\Lambda_n\bar{E}^{mn}-17iP^aS_{aim}\Lambda_n\bar{E}^{mn}+\frac{139i}{96}\slashed{P}\slashed{S}^l{}_i\gamma\cdot T^-\Lambda_l \nonumber\\
&\;\;\;-\frac{19i}{32}\gamma\cdot T^-\slashed{S}^l{}_i\slashed{P}\Lambda_l-\frac{65i}{32}\gamma\cdot T^-P^aS_a{}^l{}_i\Lambda_l+\frac{61i}{24}P^aS^{bl}{}_iT^{-ab}\Lambda_l +\frac{31i}{9}\slashed{P}E^{mn}E_{mn}\varepsilon_{ij}\Lambda^j\nonumber\\
&\;\;\;-\frac{14i}{9}\slashed{P}E^{mn}E_{im}\varepsilon_{nj}\Lambda^j+\frac{i}{2}\gamma\cdot T^-\slashed{P}\bar{E}_{ij}\Lambda^j -\frac{i}{6}\slashed{P}\gamma\cdot T^+\bar{E}_{ij}\Lambda^j-\frac{13i}{12}T^{+ab}T^-_b{}^cP_a\gamma_c\varepsilon_{ij}\Lambda^j\nonumber\\
&\;\;\;+\frac{58i}{9}\varepsilon^{kl}E_{ik}E^{mn}E_{mn}\Lambda_l -\frac{68i}{9}\varepsilon^{kl}E^{mn}E_{in}E_{mk}\Lambda_l-\frac{2i}{3}E_{im}E^{mn}\gamma\cdot T^-\Lambda_n-\frac{i}{3}\slashed{S}_{mn}E^{mn}E_{ij}\Lambda^j \nonumber\\
&\;\;\;+\frac{53i}{18}\slashed{S}_{nj}E^{mn}E_{im}\Lambda^j-\frac{19i}{6}\slashed{S}_{im}E^{mn}E_{nj}\Lambda^j+\frac{7i}{9}\slashed{S}_{ij}E^{mn}E_{mn}\Lambda^j +\frac{11i}{6}S_{aik}S^{amn}E_{mn}\varepsilon^{kl}\Lambda_l\nonumber\\
&\;\;\;-\frac{4i}{3}S^{amn}S_{amn}E_{ik}\varepsilon^{kl}\Lambda_l+\frac{19i}{6}\gamma^{ab}S_{aik}S_b^{mn}\varepsilon^{kl}E_{mn}\Lambda_l- \frac{7i}{2}\gamma^{ab}S_{amk}S_b^{mn}E_{in}\varepsilon^{kl}\Lambda_l\nonumber\\
&\;\;\;-\frac{3i}{4}\slashed{S}_{im}\gamma_{ab}S^a_{nl}S_b^{mn}\Lambda^l-\frac{i}{6}\slashed{S}_{il}S^{amn}S_{amn}\Lambda^l +\frac{i}{3} \slashed{S}_{mn}S^{amn}S_{ail}\Lambda^l+\frac{125i}{144}\gamma^a T^+_{ac}T^-{}_b{}^cS^b_{ij}\Lambda^j
\end{align}
Thereafter, taking a left-supersymmetry transformation of $\Psi_{i}$ (\ref{d4.6}) and extracting out the Lorentz invariant and SU(2)-invariant contribution $\mathcal{F}$ up to bosonic terms, we get:
\begin{align}\label{I2}
\mathcal{F}&=-16iD^2 D^{a}P_{a}-24i D^2(D)+32iD^a P_{a} D^b P_{b}-32iD^{a}P^{b}D_{(a}P_{b)}+48iD^{b}P_{b}D+24i D^2 \nonumber\\
&\;\;\; +\frac{4i}{9}C_{ijkl}C^{ijkl}-4iR(V)_{ab}{}^{i}{}_{j}R(V)^{ab}{}^{j}{}_{i}+8iR(A)_{ab}R(A)^{ab}+16iE_{ij}D^2 E^{ij}-\frac{16i}{9}D_{a}E^{ij}D_{a}E_{ij} \nonumber\\
&\;\;\; +\frac{4i}{3}D^a T^{-}_{ac}D^b T^{+}_{bc}-6i S^{a}{}^{i}{}_{j}D_{(a}D_{b)}S^{b}{}^{j}{}_{i}-4iS^{a}{}^{i}{}_{j}D^2 S_{a}{}^{j}{}_{i}-\frac{34i}{9}D_a S^{a}{}^{i}{}_{j}D_{b}S^{b}{}^{j}{}_{i}-4iD_a S_{b}{}^{i}{}_{j}D^{a}S^{b}{}^{j}{}_{i}\nonumber\\
&\;\;\; +\frac{88i}{3}E_{ij}D_{a}E^{jk}S^{a}{}^{i}{}_{k}+\frac{8i}{3}D_{a}E_{ij}E^{jk}S^{a}{}^{i}{}_{k}+\frac{154i}{9}E_{ij}E^{jk}D_{a}S^{a}{}^{i}{}_{k}+\frac{67i}{24}S^{a}{}^{i}{}_{j}S^{b}{}^{j}{}_{k}D_{a}S_{b}{}^{k}{}_{i}\nonumber\\
&\;\;\; -16i E^{ij}E_{ij}D^{a}P_{a}+4iD^{a}P_{a}S_{b}{}^{i}{}_{j}S^{b}{}^{j}{}_{i}-\frac{28i}{3}D_{a}P_{b}S^{a}{}^{i}{}_{j}S^{b}{}^{j}{}_{i}+\frac{14i}{3}D_{a}E^{ij}S_{b}{}_{ij}T^{-}{}^{ab}\nonumber\\
&\;\;\; +\frac{5i}{4}E^{ij}S_{b}{}_{ij}D_{a}T^{-}{}^{ab}-6iE^{ij}D_{a}S_{b}{}_{ij}T^{-}{}^{ab}+6iE_{ij}D_{a}S_{b}{}^{ij}T^{+}{}^{ab}+5iS_{a}{}^{i}{}_{j}S_{b}{}^{j}{}_{k}R(V)^{ab}{}^{k}{}_{i}\nonumber\\
&\;\;\; +\frac{14i}{3}R(V)^{ij}\cdot T^{+}E_{ij}-3i R(V)_{ij}\cdot T^{-}E^{ij}-\frac{i}{3}S_{a}{}^{ij}S^{a}{}^{kl}C_{ijkl}+\frac{16i}{3}C^{ijkl}E_{ij}\bar{E}_{kl}\nonumber\\
&\;\;\; -\frac{13i}{2}D S_{a}{}^{i}{}_{j}S^{a}{}^{j}{}_{i}+32iP^{b}D_{b}D^a P_a -32 iP_a D^2 P_a +48 i P^a D_a (D)-\frac{28i}{9}P^b S_{b}{}^{i}{}_{j}D^a S_{a}{}^{j}{}_{i}\nonumber\\
&\;\;\; +8iP_a S_{b}{}^{i}{}_{j}D^a S^{b}{}^{j}{}_{i}-12i P_b S_{a}{}^{i}{}_{j}D^a S^{b}{}^{j}{}_{i}+32i P^{a}P_{a}D^{b}P_{b}+64iP^a P^b D_a P_b \nonumber\\
&\;\;\; +\frac{16i}{9}P^a E_{ij}D_{a}E^{ij}+\frac{16i}{9}P^a E^{ij}D_{a}E_{ij}+\frac{4i}{3}P^aT^{-}_{ac}D_{b}T^{+}{}^{bc}+\frac{4i}{3}P^aT^{+}_{ac}D_{b}T^{-}{}^{bc}\nonumber\\
&\;\;\; +\frac{56i}{9}P^a P^b S_{a}{}^{i}{}_{j}S_{b}{}^{j}{}_{i}+\frac{4i}{3}P^a P_{a} S_{b}{}^{i}{}_{j}S^{b}{}^{j}{}_{i}-\frac{160i}{9}P^a P_{a}E_{ij}E^{ij}+\frac{4i}{3}P_{a}P_{b}T^{-}{}^{ac}T^{+}{}^{b}{}_{c}\nonumber\\
&\;\;\; +\frac{20i}{9}P^a S_{a}{}^{i}{}_{j}E_{ik}E^{jk}-\frac{41i}{12}P_a S_{b}{}_{ij}E^{ij}T^{-}{}^{ab}+ \frac{188i}{9}E^{mn}E_{mn}E^{kl}E_{kl}\nonumber\\
&\;\;\; -\frac{184i}{9}E^{mn}E_{mk}E^{kl}E_{ln}+\frac{2i}{3}T^+\cdot T^+ \bar{E}^{mn}E_{mn}+\frac{17i}{6}S_{amn}S^{akl}E_{kl}S^{mn}\nonumber\\
&\;\;\; +\frac{13i}{6}E^{mn}E_{mn}S^{akl}S_{akl}-\frac{13i}{12}\varepsilon^{abcd}S_a^{mn}S_b^{kl}S_{cmk}S_{dln} +\frac{2i}{3}S^{akl}S_{amn}S^{bmn}S_{bkl}\nonumber\\
&\;\;\; -\frac{5i}{6}S^{amn}S_{amn}S^{bkl}S_{bkl}+6iS_a{}^m{}_lS_b^{ln}T^{+ab}E_{mn}+\frac{19i}{6}T^{-ab}S_{alm}S^{bl}{}_nE^{mn}\nonumber \\
&\;\;\; -\frac{47i}{36}S^{amn}S^b_{mn}T^-_{ac}T^+{}_b{}^c\;.
\end{align}
Finally, we get the bosonic terms in the Lagrangian density $\mathcal{L}$ in (\ref{R4}) by taking $\mathcal{L}=-i\mathcal{F}+\text{h.c}$. In order to bring $\mathcal{L}$ to the form presented in (\ref{R4}), we also perform some integration by parts and extract out the bare K-gauge field $f_{a}{}^{b}$ in some terms. A non-trivial check of the above results (\ref{I}, \ref{I1}, \ref{I2}) is that they are all invariant under special conformal transformation. They also satisfy the S-transformation properties as given in (\ref{d4.9}). A further consistency check on $\mathcal{G}^{a}_{ij}$ is that it satisfies the constraint (\ref{R3}) that the embeddding of the real scalar multiplet into the abstract multiplet constituting the density formula should satisfy. In order to check the S-transformation of $\Psi_{i}$, it may be useful to have the bosonic terms of $\mathcal{D}_{ij}$ which we present below:
\begin{align}\label{I3}
\mathcal{D}_{ij}&=32iD^aD_aE_{ij}+32iD_aS^a_{m(i}E_{j)n}\varepsilon^{mn}-64i\varepsilon^{mn}D_aE_{m(i}S^a_{j)n}-\frac{64i}{3}D_aP^aE_{ij}+16iT^-_{ab}D^aS^b_{ij}\nonumber\\
&\;\;\;-8iR(V)_{ab}{}^l{}_{(i}\varepsilon_{j)l}T^{-ab}-32iDE_{ij}+\frac{16i}{3}\bar{E}^{kl}C_{ijkl}-\frac{64i}{3}P_{a}P^{a}E_{ij}+\frac{128i}{3}E^{mn}E_{mn}E_{ij}\nonumber \\
&\;\;\; -64iE^{mn}E_{m(i}E_{j)n}-4i T^-\cdot T^-\bar{E}_{ij}-8iT^{-ab}S_a{}^l{}_{(i}S_{b)jl}-\frac{40i}{3}S_{amn}S^{amn}E_{ij}\nonumber\\
&\;\;\; +16iS_a^{mn}E_{mn}S^a_{ij}
\end{align}
\section{Bianchi identity of the Bianchi identity: Constraints from Higher Bianchi follows from the preliminary constraints}\label{Bianchi}
In this section we will revisit the arguments made in \cite{Butter:2019edc} to show that the constraints coming from the higher Bianchi will follow from the preliminary constraints (\ref{d2.6}, \ref{d3.5}). The argument is based on taking a Bianchi identity of the Bianchi identity. Consider the super four form $J$ which appears in the action integral. As discussed earlier, the action integral will be supersymmetric invariant if it satisfies the Bianchi identity $I=dJ=\nabla J=0$. Now, $I$ is a super five form that is also gauge invariant under Lorentz transformations, dilations, R-symmetries, special conformal transformation as well as S-supersymmetry. Clearly $\nabla I=dI=0$ holds identically. We showed earlier that the Bianchi identity ($I$) with five gravitinos are identically zero. The non-trivial $I$ that we set to zero to get our preliminary constraints are:
\begin{align}\label{BB}
(I)_{e\bar{\psi}^3\psi}=(I)_{e\bar{\psi}\psi^3}=(I)_{e\bar{\psi}^2\psi^2}=0\;.
\end{align}
Now, let us consider $(\nabla I)_{e\bar{\psi}^4\psi}$. Since $(I)_{e\bar{\psi}^3\psi}$ is already set to zero, it can only come from $t_0 (I)_{e^2\bar{\psi}^3}$. And hence:
\begin{align}\label{BB1}
(\nabla I)_{e\bar{\psi}^4\psi}=t_0 (I)_{e^2\bar{\psi}^3}=0
\end{align}
This implies:
\begin{align}\label{BB2}
(I)_{e^2\bar{\psi}^3}=0
\end{align}
The reason for this is the following. The only way $t_0 (I)_{e^2\bar{\psi}^3}$ can be zero is either $(I)_{e^2\bar{\psi}^3}$ is $t_0$-exact or zero. But $(I)_{e^2\bar{\psi}^3}$ cannot be $t_0$ exact since it does not contain a pair of left and right handed gravitino and hence it has to be zero. Thus, we see that the Bianchi $(I)_{e^2\bar{\psi}^3}$ does not yield any new constraints but is satisfied as a consequence of the preliminary constraints.

Now, let us consider $(\nabla I)_{e\bar{\psi}^3\psi^2}$:
\begin{align}\label{BB3}
(\nabla I)_{e\bar{\psi}^3\psi^2}=t_0(I)_{e^2\bar{\psi}^2\psi}=0
\end{align}
Unlike the previous case, $(I)_{e^2\bar{\psi}^2\psi}$ can be $t_0$ exact. However, by a suitable choice of $J_{e^3 \bar{\psi}}$, we can make it zero. This is exactly how we derived our $J_{e^3 \bar{\psi}}$ (\ref{d4.1}). We can continue our reasoning along these lines to show that all the other constraints coming from the higher Bianchi identities are also automatically satisfied as a result of the preliminary constraints (\ref{d2.6}, \ref{d3.5}), and they don't yield any new constraints.

\bibliography{references}
\bibliographystyle{jhep}
\end{document}